\documentclass[a4paper,10pt]{article}
\pdfoutput=1 
\usepackage{jheppub}
\usepackage{mathrsfs}
\usepackage{amsmath}
\usepackage{amssymb}
\usepackage{subcaption}
\usepackage{amsfonts}
\usepackage[T1]{fontenc} 
\usepackage{breqn}
\usepackage{caption}
\usepackage{subcaption}
\usepackage{comment}
\excludecomment{hidden}
\usepackage{caption}
\usepackage{xcolor}
\title{\boldmath CFT Phase Transition Analysis of Charged, Rotating Black Holes in $D=4$: A Holographic Thermodynamics Approach
}

\author{Abhishek Baruah,}
\author{Prabwal Phukon}
 \affiliation{Department Of Physics,\\Dibrugarh University, Dibrugarh,786004, Assam, India}
\affiliation{Theoritical Physics Division, Center for Atmospheric Studies, Dibrugarh University}
\emailAdd{rs\_abhishekbaruah@dibru.ac.in}
\emailAdd{prabwal@dibru.ac.in}

\abstract{We investigate the holographic thermodynamics of 4-D Kerr-Newman AdS black holes, focusing on the conformal thermal states that are dual to these black holes. We explore the thermodynamic behavior within specific ensembles characterized by fixed sets of variables: $(\mathcal{Q},\mathcal{J},\mathcal{V},C)$, $(\mathcal{Q},\Omega,\mathcal{V},C)$, $(\varphi,\Omega,\mathcal{V},C)$, $(\varphi,\mathcal{J},\mathcal{V},C)$, $(\mathcal{Q},\Omega,p,C)$, and $(\varphi,\Omega,p,C)$. Here, $\varphi$, $\mathcal{Q}$, $\Omega$, $\mathcal{J}$, $p$, $\mathcal{V}$, and $C$ represent the electric potential, electric charge, angular velocity, angular momentum, CFT pressure, CFT volume, and central charge, respectively. The inclusion of both charge and momentum significantly enriches the regime of phase transitions, leading to a variety of phenomena including first-order Van der Waals-type phase transitions, (de)confinement phase transitions, Davies-type phase transitions, and second-order superfluid $\lambda$-type phase transitions. Notably, the introduction of the CFT pressure variable allows us to identify phase transitions and critical behavior in the $(\mathcal{Q},\Omega,p,C)$ and $(\varphi,\Omega,p,C)$ ensembles, which had not been previously observed. This study underscores the complexity and richness of phase transitions in these systems due to the inclusion of both charge and angular momentum.

}

\begin{document} 
\maketitle
\flushbottom

\section{Introduction}\label{sec:1}
Understanding the underlying similarities between the laws of black hole mechanics and thermodynamics about half a century back culminated in the formulation of what is now known as black hole thermodynamics \cite{a1,a2,a3, a,b,a4,b1,b2}. The two seminal equations of black hole thermodynamics given by Hawking and Bekenstein established black holes not merely as astrophysical objects but as thermodynamic systems characterized by entropy and temperature. These are as follows:- 
\begin{equation}
\label{eq:ETS}
T=\frac{\kappa}{2\pi}, \quad S=\frac{A}{4 G_N}
\end{equation}

Here $\kappa$, $A$ and $G_N$ are surface gravity, horizon area, and Newton's universal gravitational constant respectively.  
One particular class of black holes that have contributed immensely to the subsequent development and expansion of conventional black hole thermodynamics is the asymptotically anti-de-Sitter (AdS) black holes. AdS black holes, depending on their nature (uncharged, charged, rotating, hairy) and the ensemble under consideration, are characterized by a diverse range of rich phase structures \cite{b3,b4,b5,j,d,k}. At present, apart from the traditional thermodynamic methods, several alternative tools like thermodynamic geometry \cite{af1,af2,af3,af4,af5,af6,af7,af8} and thermodynamic topology \cite{g1,g2,g3,g4,g5,g6} are used to gain useful insights into these phase structures.\\
From the traditional black hole thermodynamics, next comes the Extended Phase Space Thermodynamics (EPST). This concept was first given by Kastor, Ray, Traschen, and later more insights and development were done by other researchers. This EPST suggested the introduction of a new pair of thermodynamic variables, namely the negative cosmological constant which is linked to the thermodynamic pressure, treating it as a state variable and its conjugate variable, namely the thermodynamic volume \cite{f,g,h,i}. The thermodynamic pressure is given as:-
\begin{equation}
P=-\frac{\Lambda}{8\pi G_N}, \quad \quad \Lambda=-\frac{d(d-1)}{2L^2}
\end{equation} 
where $L$ is given as the AdS curvature radius, $G_N$ is the Newton's constant,  $D=d+1$ denotes the number of bulk spacetime dimensions and $V=-\Theta$ is identified as the thermodynamic volume in the extended first law of black holes.
But in this formalism, a slightly odd way to introduce the mass parameter as the enthalpy rather than the internal energy.\\
From the holographic perspective, understanding the dimensional-dependent factors and the cosmological constant $\Lambda$ term in the generalized Smarr relation presents significant challenges. The AdS/CFT correspondence, also known as gauge/gravity duality, posits that the thermodynamics of black holes within the "bulk" spacetime is mirrored by the thermodynamics of large-N strongly coupled gauge theories on the asymptotic boundary of this spacetime \cite{c,c1}.
One of the key phenomena that exemplifies this duality is the Hawking-Page transition observed between thermal radiation and a large AdS black hole, which corresponds to the (de)confinement phase transition in the gauge field theories. In this framework, the thermodynamic variables for black holes, such as entropy $S$, and temperature $T$ as described by the equation \eqref{eq:ETS}, should correlate directly with the thermal states of the boundary field theory. The generalized Smarr relation, which connects these variables in gravitational theory, is expected to have a one-to-one correspondence with the Euler relation in the field theory thermodynamics.
However, the interpretation of the cosmological constant $\Lambda$ as pressure in the bulk Smarr formula does not straightforwardly translate to the field theory context. Consequently, while the bulk pressure $P$ is conceptually analogous to the boundary CFT pressure $p$, they do not match quantitatively. Similarly, the thermodynamic volume in the bulk does not correspond to the spatial volume $\mathcal{V}$ of the CFT at the boundary. This disconnection arises from the different roles and interpretations of $\Lambda$ and the volume terms in their respective frameworks.
The challenge stems from reconciling the intrinsic thermodynamic interpretations in the AdS/CFT duality, where bulk parameters like $\Lambda$ have direct thermodynamic significance, with their boundary counterparts, which lack a straightforward thermodynamic interpretation. This discrepancy indicates a deeper and more complex relationship between the bulk and boundary descriptions, suggesting that the dual field theory may capture more than just the conventional thermodynamic variables, possibly encoding information about the underlying quantum structure of spacetime itself. \\
 The study of thermal equilibrium of AdS black holes has achieved a variety of interesting behaviors such as the first order phase transition to the radiation which is analogous to the confinement/deconfinement of the quark-gluon plasma \cite{e}, also for charged AdS black that undergoes a Van der Waals type phase transition \cite{l,m} and also for rotating black holes \cite{n,o,p}. The study of black hole thermodynamics \cite{q} garnered study in multiple fields such as polymer transitions \cite{m1}, superfluid transitions \cite{r}, triple points \cite{r1,r2} and reentrant phase transitions \cite{s,s1}, micro-structures of black hole \cite{t} and multi-critical phase transitions \cite{u,v}. In a recent study, Visser \cite{bb} gave a set of thermodynamic variables to this formalism that is referred to as central charge $C$ and the chemical potential $\mu$. With this extension, the study of the central charge criticality has come to existence \cite{cc1}.\\
The more potent phase behavior of the AdS black hole is due to the incorporation of the negative cosmological constant $\Lambda$, as an additional thermodynamic variable leading to the existence of new phase transitions. Also, the variation of Newton's constant has been added to the EPST formalism \cite{w,aa1}. Newton's constant is used as a "book-keeping" device for finding the correct thermodynamical interpretation for the black holes extended first law. The extended first law and the generalized Smarr relation in $D=d+1$ dimensions can be given as:-
\begin{equation}
dM=T\delta S+\Phi\delta Q+\sum_i\Omega_i \delta J_i+V \delta P, \quad
M=\frac{d-1}{d-2}\left(TS+\sum_i \Omega_i J_i\right)+\Phi Q-\frac{2}{d-2}PV
\end{equation}
where $T$ is the Hawking temperature, $Q$ is the electric charge, $\Phi$ is the electric potential, $S$ is the entropy, $J_i$ is the respective angular momenta and the corresponding angular velocities between horizon and infinite is $\Omega_i$.
In recent years the study of holographic explanation of black holes has been studied intensively \cite{x,x1,y,z}.
In many works, it is said that when we vary $\Lambda$, it is related to varying the number of colors $N$, or the number of degrees of freedom $N^2$, in the boundary field theory. Varying the $N$ will in turn correspond to vary the number of branes for the gauge theories arising from the coincident D-branes. Following in the conformal field theories (CFTs), the central charge $C$ is given by the number of degrees of freedom, which when varied takes us from one CFT to another. The holographic CFT which also dual to the Einstien gravity, the central charge $C$ corresponds to $C \sim L^{d-1}/G$, where $G$ is Newton's constant in $d+1$ dimensions and $L$ is the radius of curvature of the bulk geometry, which relates to $\Lambda$ by $\Lambda=-d(d-1)/2L^2$. Therefore, varying $G$ and $\Lambda$ in the bulk theory is dual to varying $C$ in the boundary CFT. In  \cite{aa} it is seen that when we vary $\Lambda$ in the bulk CFT not only corresponds to a change in the $C$ (or $N^2$), but it also suggests a change in the volume $\mathcal{V}$ of the spatial boundary. The reason is that the boundary curvature radius of a particular boundary metric is equal to the bulk curvature radius $L$. We also see that the electric potential and its analogous CFT electric charge are related to the AdS length scale. Hence we cannot directly project the first law thermodynamics of the bulk spacetime into the holographic dual boundary theory. To solve this projection dissimilarity issue the researchers have given out a theory to maintain the CFT's central charge fixed by dynamically varying the CFT volume $\mathcal{V}$ and the CFT central charge $C$ without introducing the variable Newton's constant \cite{bb1,bb2,bb3,cc}. The bulk AdS space-time conformal completion is given as:-
 \begin{equation}
 ds^2=\omega(-dt^2+l^2d\Omega^2_{d-1}) \end{equation}
where the dimensionless arbitrary conformal factor is $\omega$ which reflects the conformal symmetry of the boundary theory. $d\Omega^2_{d-1}$ is considered as a metric on a unit $(d-1)$ dimensional sphere having the corresponding volume as $\Omega_{d-1}$. By taking $\omega$ as an independent coordinate, the CFT volume reads as:-
\begin{equation}
\mathcal{V}=\Omega_{d-1}R^{d-1}
\end{equation}
 where $R=\omega L$ is given as the curvature radius which is considered a variable on the plane where the CFT resides. The expression of the central charge $C$ to which the CFT volume $\mathcal{V}$ is independent is given as:-
 \begin{equation}
 \label{eq:central}
 C=\frac{\Omega_{d-1} L^{d-1}}{16\pi G_N}
 \end{equation}
 In various dimensions of the CFT, there are several nominees for the central charge. If we insert $d=2$ the usual central charge will equate to our central charge as $C=c/12$ and correspondingly inserting it in \eqref{eq:central} we get the famous Brown-Henneaux dictionary which relates in $AdS_3/CFT_2$ as $c=3L/2G_N$ \cite{ce1}. Also there are other nominees for the central charge which are denoted as $C_T$ and $a^*_d$ which is seems to be defined in odd dimensions \cite{ce2,ce3}. The $C_T$ which is taken as the CFT stress tensor function is particularly defined as the total normalization \cite{ce4} and the $a^*_d$ is taken for the spherical-shaped regions as the coefficient for the vacuum entanglement entropy \cite{ce5}. Using the AdS CFT holographic dictionary:- 
 \begin{equation}
 E=\frac{M}{\omega},\quad \mathcal{T}=\frac{T}{\omega},\quad \mathcal{S}=S, \quad \mathcal{J}=J, \quad \varphi=\frac{\Phi \sqrt{G_N}}{\omega L},\quad \mathcal{Q}=\frac{Q L }{\sqrt{G_N}}
 \end{equation}
 Using the dictionary given above the bulk first law is given as:-
 \begin{equation}
 \delta E=\mathcal{T}\delta \mathcal{S}-\varphi \delta \mathcal{Q}+\mu \delta C+\Omega \delta \mathcal{J}-p \delta \mathcal{V}
 \end{equation}
 and the Smarr relation is given as:-
\begin{equation}
E=\mathcal{T}\mathcal{S}+\varphi \mathcal{Q}+\mu C+\Omega J
\end{equation}
followed by the two expressions for the chemical potential $\mu$ and for the CFT pressure $p$ respectively is given as:-
\begin{equation}
\mu=\frac{1}{C}(E-\mathcal{T}\mathcal{S}-\Omega J-\varphi Q), \quad p=\frac{E}{(d-2)\mathcal{V}}
\end{equation}
The analogous dual CFT does not change even if we see rescaling in the bulk cosmological constant. As a consequence, we can calculate the first law of the black hole thermodynamics, from the dual boundary field theory. In this way, it explains a holographic explanation of the black hole chemistry, ideally for the Reissner-Nordstrom AdS black holes Van der Waals phase transition. This phase change is determined by the analogous field theory degrees of freedom, which agrees with the central charge criticality examinations for the Gauss-Bonnet black holes \cite{gg, hh} and for non-linear electromagnetic black holes \cite{ee,ff}. In the analogous CFT description, these black holes exhibit zeroth-order, first-ordered, and second-ordered phase transitions but upon careful examination, it was revealed that there is no criticality in the pressure-volume CFT, which verifies that the CFT states which is regarded as a dual to the charged AdS black holes in not a Van der Waals fluid.\\
The main aim of this paper is to further generalize or we can say incorporate together previous examinations of the holographic thermodynamics namely the CFT thermodynamics of charged \cite{dd} and the CFT thermodynamics of rotating black holes \cite{ii,dd1}. We can divide it into two main ideas. Firstly, we review the previously studied generalized mass/energy formulas for the charged and rotating Kerr-Newman AdS black holes (KN-AdS) from both sides namely the holographic CFT side and the bulk AdS. Secondly, taking the ensembles from the CFT side, we investigate the ensembles namely the 6 ensembles, and study in detail of each ensemble varying all the parameters. In doing so we can examine if some ensembles do exhibit phase transitions and exhibit criticality in the CFT thermodynamics. \\
The motivation for this work is to study the phase transitions and critical phenomena in charged rotating AdS black holes within the framework of holographic thermodynamics.  Building on the recent study in \cite{dd}, which explored the phase transitions of charged AdS black holes and uncovered new ensembles not seen in traditional black hole thermodynamics, we aim to extend this framework to investigate the holographic thermodynamics of rotating charged black holes. While charge ($\mathcal{Q}$) and angular momentum ($\mathcal{J}$) are inherent in the metric of charged rotating AdS black holes, this study focuses on examining the effects of introducing $\mathcal{V}$ (CFT volume), $p$ (CFT pressure) and $C$ (central charge) as additional parameters in the thermodynamic phase space, which arise naturally in the context of holographic thermodynamics. By analyzing specific ensembles—fixed $(\mathcal{Q}, \mathcal{J}, \mathcal{V}, C)$, $(\mathcal{Q}, \Omega, \mathcal{V}, C)$, $(\varphi, \Omega, \mathcal{V}, C)$, $(\varphi, \mathcal{J}, \mathcal{V}, C)$, $(\mathcal{Q}, \Omega, p, C)$, and $(\varphi, \Omega, p, C)$—we uncover diverse phase transition phenomena, including first-order Van der Waals-type transitions, Davies-type transitions, and second-order superfluid $\lambda$-type transitions. Notably, the introduction of this framework and these previously unexplored ensembles has led to the identification of phase transitions that were absent in traditional thermodynamics or the extended phase space thermodynamic (EPST) formalism. The inclusion of $p$ and $C$ enables the discovery of novel critical behaviors, particularly in the $(\mathcal{Q}, \Omega, p, C)$ and $(\varphi, \Omega, p, C)$ ensembles, which were not previously examined. \\
This paper is divided into 5 sections. In section \ref{sec:1} we have given the introduction of the paper as well as the approach as to what we want to study and also the motivations. In section \ref{sec:2}, we review the mass-energy formulas from the extended phase space to the CFT thermodynamics. In section \ref{sec:3}, we have the main results where study extensively the criticality or phase transitions for various ensembles for the thermal states analogous to the 4-D KN-AdS black hole. In section \ref{sec:5} we give out the conclusion.
\section{Mass energy formulas from the Eextended Thermodynamics to the CFT Thermodynamics}
\label{sec:2}
We review the mass formulas for the charged and rotating black holes \cite{ii}. We do this in the Extended Phase Space Thermodynamics in terms of the bulk event horizon area, electric charge, angular momentum, and the cosmological constant. Then after this, we will look into the black hole mass formula in the so-called "mixed thermodynamics" where there is a dependence on black hole entropy, electric charge, angular momentum, thermodynamic pressure, and the boundary CFT central charge. After all this, we review the energy formulas of the dual CFT theory from the mass formulas of the Extended Phase Space Thermodynamics of the KN-AdS black holes.
\subsection{Extended Phase Space Thermodynamics (EPST)} 
The Einstein-Maxwell theory action is taken here as:-
 \begin{equation}
 I=\frac{1}{16 \pi G_N} \int d^4x\sqrt{-g}(R-2\Lambda-\mathcal{F}^2)
 \end{equation}
 Here, the $\Lambda$ is regarded as the cosmological constant, $G_N$ as the gravitational constant and $\mathcal{F}_{ab}$ as the strength of the $U(1)$ field, and lastly $R$ is the Ricci scalar. The solution is given below:-
 \begin{equation}
 ds^2=-\frac{\Delta_r}{\Sigma^2}\left(dt-\frac{a \sin^2\theta}{\Xi}d\phi\right)^2+\sigma^2\left(\frac{dr^2}{\Delta_r}+\frac{d\theta^2}{H_\theta}\right)+\frac{H_\theta \sin^2\theta}{\Sigma^2}\left[adt-\frac{r^2+a^2}{\Xi}d\phi \right]^2
\end{equation} 
\begin{equation}
\mathcal{F}_{ab}=(dB)_{ab}, \quad
B=-\frac{G_Nqr}{\Xi^2}\left[dt-a\sin^2\theta d\phi \right]+\Phi_tdt
\end{equation}
where all the unknown terms are:-
\begin{equation}
\begin{split}
&\Delta_r=r^2-2G_Nmr+a^2+G_N^2q^2+\frac{(r^2+a^2)r^2}{l^2}\\
& H_\theta=1-\frac{a^2}{l^2}\cos^2\theta, \quad \Sigma^2=r^2+a^2\cos^2\theta, \quad \Phi_t=\frac{G_Nqr_+}{a^2+r^2_+},\quad \Xi=1-\frac{a^2}{l^2}.
\end{split}
\end{equation}
From the black hole when we set $\Delta_r=0$ we get the event horizon $r_+$. Here, the gauge potential is given by $B_a$, where we get vanishing gauge potential for the black holes for the values of $\Phi_t$. The related parameters namely the ADM mass $m$, electric charge $q$ and the angular momentum $J$ are given as:-
\begin{equation}
M=\frac{m}{\Xi^2}, \quad
J=\frac{am}{\Xi^2}=Ma,\quad Q=\frac{q}{\Xi}
\end{equation}
The cosmological constant is given ($d=3$) as:-
\begin{equation}
\label{eq:lambda}
\Lambda=-\frac{3}{L^2}
\end{equation}
where the curvature radius or AdS length scale is given as $L$.
The horizon area $A$, thermodynamic pressure $P$ are:-
\begin{equation}
\label{eq:pressure}
 A=\frac{4\pi(r^2_++a^2)}{\Xi}, \quad
 P=-\frac{\Lambda}{8\pi G_N}
\end{equation}
The temperature, angular velocity thermodynamic volume, and electric potential are given as
\begin{equation}
T=\frac{-L^2(a^2+G^2_Nq^2)+r^2_+(a^2+L^2)+3r^2_+}{4\pi L^2r_+(a^2+r^2_+)}, \quad
\Omega_b=\frac{a(L^2+r^2_+)}{L^2(a^2+r^2_+)}
\end{equation}
\begin{equation}
V=\frac{2\pi L^2(a^2L^2(a^2+G^2_Nq^2)-(r^4_+(a^2-2L^2))-r^2_+(a^4-3a^2L^2))}{3r_+(a^2-L^2)^2}, \quad
\Phi=\frac{G_Nqr_+}{a^2+r^2_+}
\end{equation}
We can relate the surface gravity $\kappa$ and the entropy $S$ from the area $A$ and the temperature $T$ as:-
\begin{equation}
\label{eq:entropy}
S=\frac{A}{4G_N}=-\frac{\pi  L^2 \left(a^2+r^2\right)}{G (a-L) (a+L)}, \quad
\kappa=2\pi T
\end{equation}
The KN- AdS black hole generalised mass $M$ in terms of $(G_N,Q,J,P,S)$ is given as
\begin{equation}
\label{eq:mass}
M^2=\frac{12 \pi ^2 J^2 \left(8 G^2 P S+3\right)+16 G^3 P S^2 \left(4 P S^2+3 \pi  Q^2\right)+G^2 \left(48 P S^3+9 \pi ^2 Q^4\right)+18 \pi  G Q^2 S+9 S^2}{36 \pi  G S}
\end{equation}
The thermodynamic first law of the black hole  by using the above mass formula and Smarr relation in the EPST is given by putting $D=d+1=3+1$ as:-
\begin{equation}
\label{eq:first law}
\delta M=T\delta S+\Phi \delta Q+V\delta P+\Omega\delta J, \quad
M=2TS+2\Omega J+\Phi Q-2PV
\end{equation}
We can also write the mass formula by replacing the entropy $S$ with the area $A$ \eqref{eq:entropy} and the pressure $P$ with cosmological constant $\Lambda$ \eqref{eq:pressure}. If we wish to incorporate the dynamic nature of Newton's gravitation constant $G_N$ \cite{jj}. The new mass formula looks like:-
\begin{equation}
M^2=\frac{A^3 \Lambda ^2}{2304 \pi ^3 G^3}-\frac{A^2 \Lambda }{96 \pi ^2 G^2}+\frac{\pi  \left(G^2 Q^4+4 J^2\right)}{A}+\frac{A \left(3-2 G^2 \Lambda  Q^2\right)}{48 \pi  G^2}+\frac{1}{6} \left(3 Q^2-2 J^2 \Lambda \right)
\end{equation}
Then the first law of thermodynamics and Smarr relation treating $\Lambda$, $G_N$ as parameters as:-
\begin{equation}
\label{eq:firstlaw}
\delta M=\frac{\kappa}{8\pi G_N}\delta A+\Phi \delta Q+\Omega\delta J-\frac{V}{8\pi G_N}\delta \Lambda-(M-\Phi Q-\Omega_b J)\frac{\delta G_N}{G_N}, \quad M=\frac{\kappa A}{4 \pi G_N}+2\Omega J+\Phi Q+\frac{V \Lambda}{4 \pi G_N}
\end{equation}
We can derive the related quantities as:-
\begin{equation}
\frac{\kappa}{8\pi G_N}=\left(\frac{\partial M}{\partial A}\right)_{Q,J,\Lambda,G_N}=
\frac{16 \pi ^2 A^2 G_N \left(3-2 G_N^2 \Lambda  Q^2\right)-16 \pi  A^3 G_N \Lambda +A^4 \Lambda ^2-768 \pi ^4 G_N^3 \left(G_N^2 Q^4+4 J^2\right)}{32 \pi ^{3/2} A^2 G_N^3 A_1}
\end{equation}

\begin{equation}
\Phi=\left(\frac{\partial M}{\partial Q}\right)_{A,J,\Lambda,G_N}=
\frac{2 \left(-A^2 \Lambda  Q+12 \pi  A Q+48 \pi ^2 G_N^2 Q^3\right)}{A A_1}
\end{equation}

\begin{equation}
-\frac{V}{8\pi G_N}=\left(\frac{\partial M}{\partial \Lambda}\right)_{A,Q,J,G_N}=
\frac{-12 \pi  A^2 G_N+A^3 \Lambda -48 \pi ^2 A G_N^3 Q^2-384 \pi ^3 G_N^3 J^2}{48 \pi ^{3/2} G_N^3 A_1}
\end{equation}

\begin{equation}
\frac{\Omega J+\Phi Q-M}{G_N}=\left(\frac{\partial M}{\partial G_N}\right)_{A,Q,J,\Lambda}=\frac{16 \pi  A^3 G_N \Lambda -96 \pi ^2 A^2 G_N-A^4 \Lambda ^2+1536 \pi ^4 G_N^5 Q^4}{32 \pi ^{3/2} A G_N^4 A_1}
\end{equation}
Where the value of $A_1$ is in Appendix \ref{App}.
\subsection{Mixed Thermodynamics}
To incorporate the boundary central charge from the thermodynamic first law, we take into consideration the holographic dual relation between the AdS scale $L$, Newton's constant $G_N$, and the central charge $C$ from \eqref{eq:central} putting $d=3$ which is given as:- 
\begin{equation}
\label{eq:central}
C=\frac{\Omega_2l^2}{16 \pi G_N}
\end{equation} 
Here, $\Omega_2=4\pi$.
In terms of the bulk thermodynamic pressure and the central charge, Newton's constant $G_N$ can be achieved by combining \eqref{eq:pressure}, \eqref{eq:lambda} as:-
\begin{equation}
\label{eq:G}
G_N=\frac{1}{4}\sqrt{\frac{3}{2\pi C P}}
\end{equation}
Now of we incorporate the value of Newton's gravitational constant from the above equation \eqref{eq:G} to the mass formula in \eqref{eq:mass} we get:-
\begin{equation}
\label{eq:mass1}
\begin{split}
&M^2=\frac{\left(24 \pi ^{3/2} C P \left(S \left(\sqrt{6 \pi } Q^2 \sqrt{\frac{1}{C P}}+4 S\right)+16 \pi ^2 J^2\right)+\pi ^{5/2} \left(96 J^2 P S+9 Q^4\right)\right)}{\sqrt{CP} \sqrt{6} \pi ^2 S}\\
&+\frac{\left(8 \sqrt{6} P^2 S^4 \sqrt{\frac{1}{C P}}+6 \pi  \sqrt{6} P Q^2 S^2 \sqrt{\frac{1}{C P}}+48 \sqrt{\pi } P S^3\right)}{\sqrt{CP} \sqrt{6} \pi ^2 S}
\end{split}
\end{equation}
The thermodynamic first law in bulk-boundary corresponding to the above mass formula  \eqref{eq:mass1} is:-
\begin{equation}
dM=TdS+\Phi dQ+V_e dP+\mu_edC+\Omega dJ
\end{equation}
The pressure $P$ has a conjugate quantity the effective thermodynamic volume $V_e$ and the central charge $C$ conjugate quantity is a chemical potential $\mu_e$ where the expression for the terms is given below:-
\begin{equation}
V_e=V+\frac{\beta}{2 P}-\frac{TS+VP}{2}, \quad
 \mu_e=\frac{\beta}{C}-\frac{TS+VP}{2C}
\end{equation}
where $\beta=M-\Omega J-\frac{Q \phi}{2}$. One can refer to \cite{gg,fir,fir1} for further idea on mixed or bulk boundary thermodynamics. 
 The related quantities can be calculated as given below:-
\begin{equation}
\begin{split}
&T=\left(\frac{\partial M}{\partial S}\right)_{Q,P,C,J}\\=
&\frac{ \left(-32 \pi ^{3/2} C P \left(4 \pi ^2 J^2-S^2\right)+8 \sqrt{6} P^2 S^4 \sqrt{\frac{1}{C P}}+2 \pi  \sqrt{6} P Q^2 S^2 \sqrt{\frac{1}{C P}}+32 \sqrt{\pi } P S^3-3 \pi ^{5/2} Q^4\right)}{\sqrt[4]{\frac{2}{3}}\sqrt{CP} 8 \pi  S^{3/2} A_2}
\end{split}
\end{equation}
\begin{equation}
\Phi=\left(\frac{\partial M}{\partial Q}\right)_{S,P,C,J}=
\frac{\sqrt[4]{3} Q \left(\pi  C \left(3 \sqrt{2 \pi } Q^2 \sqrt{\frac{1}{C P}}+8 \sqrt{3} S\right)+2 \sqrt{3} S^2\right)}{2\ 2^{3/4} C S^{1/2} A_2}
\end{equation}
\begin{equation}
\begin{split}
&V_C=\left(\frac{\partial M}{\partial P}\right)_{S,Q,J,C}\\=
&\frac{C \left(96 \pi ^{3/2} C P \left(4 \pi ^2 J^2+S^2\right)+16 \sqrt{6} P^2 S^4 \sqrt{\frac{1}{C P}}+\pi ^{5/2} \left(96 J^2 P S-9 Q^4\right)+48 \sqrt{\pi } P S^3\right)}{\left(CP\right)^{3/2}16 \sqrt[4]{2} 3^{3/4} \pi  S^{1/2} A_2}
\end{split}
\end{equation} 
\begin{equation}
\resizebox{1\linewidth}{!}{$
\begin{split}
&\mu=\left(\frac{\partial M}{\partial C}\right)_{S,Q,P,J}\\=
&-\frac{P \left(-96 \pi ^{3/2} C P \left(4 \pi ^2 J^2+S^2\right)+16 \sqrt{6} P^2 S^4 \sqrt{\frac{1}{C P}}+12 \pi  \sqrt{6} P Q^2 S^2 \sqrt{\frac{1}{C P}}+\pi ^{5/2} \left(96 J^2 P S+9 Q^4\right)+48 \sqrt{\pi } P S^3\right)}{\left(CP\right)^{3/2}16 \sqrt[4]{2} 3^{3/4} \pi  S^{1/2} A_2}
\end{split}
$}
\end{equation}
\begin{equation}
\Omega_b=\left(\frac{\partial M}{\partial J}\right)_{S,Q,P,C}=
\frac{4\ 2^{3/4} \sqrt[4]{3} \pi ^{3/2} J P \sqrt{\frac{1}{C P}} (4 \pi  C+S)}{S^{1/2} A_2}
\end{equation}
Where $A_2$ is in Appendix \ref{App}. The thermodynamics of Kerr-Ads black holes have been studied in depth in \cite{kk,ll,mm}. 
\subsection{CFT Thermodynamics}
In this section, let us now review the mass of the CFT thermodynamics for the single-charged and rotating black hole. The metric is set as \cite{nn,oo,pp,qq}:-
\begin{equation}
ds^2=\omega^2(-dt^2+L^2d\Omega^2_2)
\end{equation}
The two-sphere line element is given as $d\Omega^2_2$ and $\omega$ is an arbitrary dimensionless conformal factor. In some papers, we have noticed that they have taken the conformal factor $\omega=\frac{R}{l}$ where $R$ is the curvature of the radius. But by taking into consideration paper \cite{pp} and allowing it to be a general parameter, the parameter $\omega$ allows us to examine a holographic first law keeping Newton's constant fixed and a varying cosmological constant $\Lambda$.
The boundary sphere's spatial volume is given as:-
\begin{equation}
\label{eq:volume}
\mathcal{V}=\Omega_2(\omega L)^2
\end{equation}
For this volume $\mathcal{V}$ which is given in the CFT context, an analogous pressure $p$ is also there so a work term $-pd\mathcal{V}$ exists. By using the holographic dictionary in the bulk mass, entropy, temperature, electric potential, angular momentum, and charge along with their analogous are given as \cite{j,bb,qq}:-
\begin{equation}
\label{eq:holographic}
E=\frac{M}{\omega}, \quad \mathcal{T}=\frac{T}{\omega},\quad
 \mathcal{S}=S, \quad \mathcal{J}=J, \quad \varphi=\frac{\Phi}{\omega l}, \quad \mathcal{Q}=Q l.
\end{equation}
By using \eqref{eq:pressure}, \eqref{eq:central}, \eqref{eq:holographic} we get the CFT energy as:-
\begin{equation}
\label{eq:energy}
E^2=\frac{64 \pi ^4 C^2 \mathcal{J}^2+16 \pi ^2 C^2 \mathcal{S}^2+16 \pi ^3 C \mathcal{J}^2 \mathcal{S}+8 \pi ^3 C \mathcal{Q}^2 \mathcal{S}+8 \pi  C \mathcal{S}^3+2 \pi ^2 \mathcal{Q}^2 \mathcal{S}^2+\pi ^4 \mathcal{Q}^4+\mathcal{S}^4}{4 \pi ^2 C \mathcal{S} \mathcal{V}}
\end{equation}
The thermodynamic first law and the Smarr relation are related as:-
\begin{equation}
\label{eq:first law 1}
dE=\mathcal{T}d\mathcal{S}+\varphi d\mathcal{Q}-\ pd\mathcal{V}+\mu dC+\Omega d\mathcal{J}, \quad
E=\mathcal{T}\mathcal{S}+\varphi\mathcal{Q}+\mu C+\Omega \mathcal{J}
\end{equation}
Now, calculating the related quantities we can write them as:-
\begin{equation}
\mathcal{T}=\left(\frac{\partial E}{\partial \mathcal{S}}\right)_{\mathcal{Q},\mathcal{V},C}=\frac{16 \pi ^2 C^2 \left(\mathcal{S}^2-4 \pi ^2 \mathcal{J}^2\right)+16 \pi  C \mathcal{S}^3+2 \pi ^2 \mathcal{Q}^2 \mathcal{S}^2-\pi ^4 \mathcal{Q}^4+3 \mathcal{S}^4}{4 \pi \mathcal{S} \sqrt{C \mathcal{V}} A_3}
\end{equation}
\begin{equation}
\varphi=\left(\frac{\partial E}{\partial \mathcal{Q}}\right)_{\mathcal{S},\mathcal{V},C}=\frac{\pi  \mathcal{Q} \left(4 \pi  C \mathcal{S}+\pi ^2 \mathcal{Q}^2+\mathcal{S}^2\right)}{\sqrt{C \mathcal{S} \mathcal{V}} A_3}
\end{equation}
\begin{equation}
\label{eq:eos}
-p=\left(\frac{\partial E}{\partial \mathcal{V}}\right)_{\mathcal{S},\mathcal{Q},C}=-\frac{A_3}{4 \pi  \mathcal{V}^{3/2}\sqrt{CS}}
\end{equation}
\begin{equation}
\label{eq:mu}
\mu =\left(\frac{\partial E}{\partial C}\right)_{\mathcal{S},\mathcal{Q},\mathcal{V}}=\frac{16 \pi ^2 C^2 \left(4 \pi ^2 \mathcal{J}^2+\mathcal{S}^2\right)-\left(\pi ^2 \mathcal{Q}^2+\mathcal{S}^2\right)^2}{4 \pi  C^{3/2} \sqrt{\mathcal{S} \mathcal{V}}}
\end{equation}
\begin{equation}
\Omega=\left(\frac{\partial E}{\partial \mathcal{J}}\right)_{\mathcal{S},\mathcal{Q}, \mathcal{V}}=\frac{8 \pi ^2 \mathcal{J} (4 \pi  C+\mathcal{S})}{C^{-1/2}\sqrt{\mathcal{S} \mathcal{V}} A_3}
\end{equation}
Here \eqref{eq:eos} is the CFT equation of state. The conjugate for the central charge is the chemical potential $\mu$ \eqref{eq:mu} and the value of $A_3$ is in Appendix \ref{App}.
From the bulk mass formula \eqref{eq:mass}, we see that the internal energy formula \eqref{eq:energy} can be formulated. Also, we can relate that the bulk thermodynamics first law \eqref{eq:first law} corresponds to the conformal field theory (CFT) first law \eqref{eq:first law 1}, where we note that the appearance of the variable Newton's constant is not there on incorporating a variable cosmological constant $\Lambda$.\\

\section{List of thermodynamic ensembles in the CFT}
\label{sec:3}
In recent papers \cite{dd}, the CFT states phase transitions that relate to the charged AdS black hole were examined. Here we try to add one more variable namely $\mathcal{J}$ and its conjugate $\Omega$ for the rotation to make it a charge and rotating black holes which we will only look into the corresponding CFT states of the KN-AdS black holes. Even though there are 16 ensembles to examine, it is quite challenging to study all the ensembles so we only took 6 ensembles where we see interesting phase transitions and criticality. The ensembles are fixed $(\mathcal{Q},\mathcal{J},\mathcal{V},C)$, $(\mathcal{Q},\Omega,\mathcal{V},C)$, $(\varphi,\Omega,\mathcal{V},C)$, $ (\varphi,\mathcal{J},\mathcal{V},C)$, $(\mathcal{Q},\Omega,p,C)$, $(\varphi,\Omega,p,C)$  ensembles respectively. 
\subsection{Ensemble at fixed $(\mathcal{Q},\mathcal{J},\mathcal{V},C)$}
\label{sec:en1}
\begin{figure}[htp]
\centering
\begin{subfigure}[b]{0.3\textwidth} 
\includegraphics[width=1\textwidth]{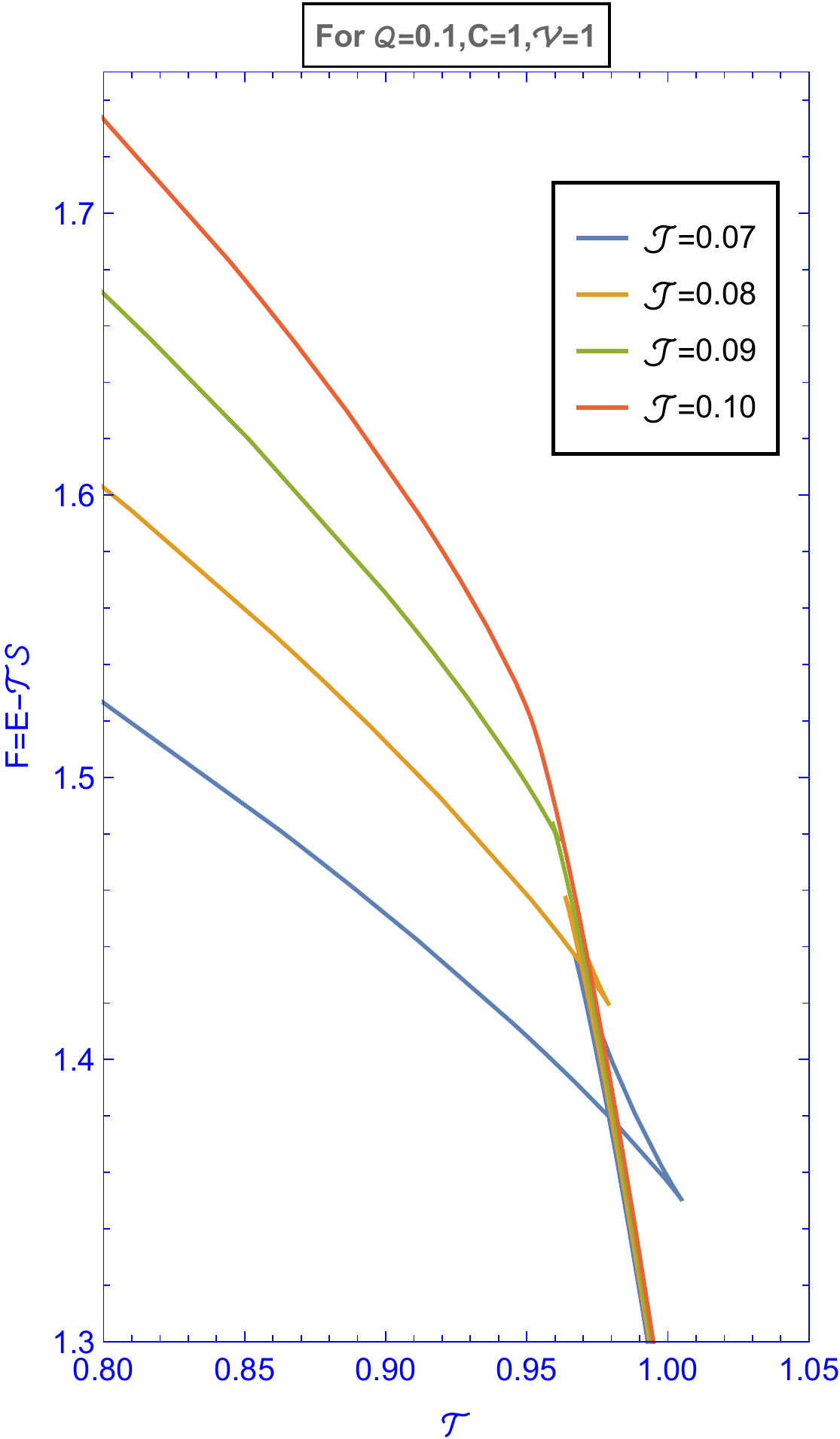}
\caption{}
\label{fig:plot1}
\end{subfigure}
\begin{subfigure}[b]{0.3\textwidth}
\includegraphics[width=1\textwidth]{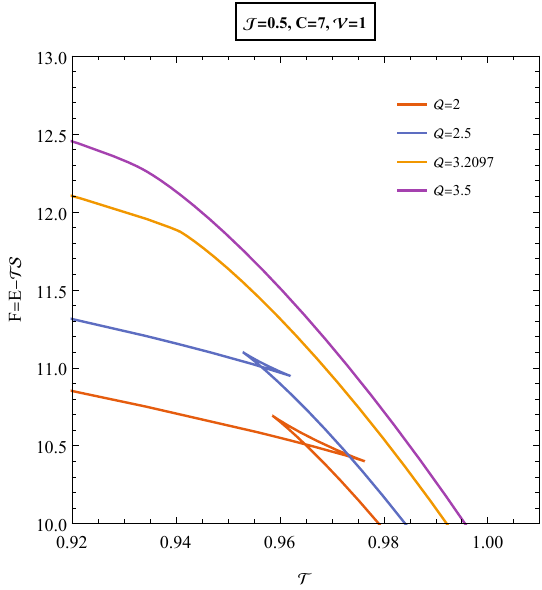}
\caption{}
\label{fig:plot2}
\end{subfigure}
\begin{subfigure}[b]{0.3\textwidth} 
\includegraphics[width=1\textwidth]{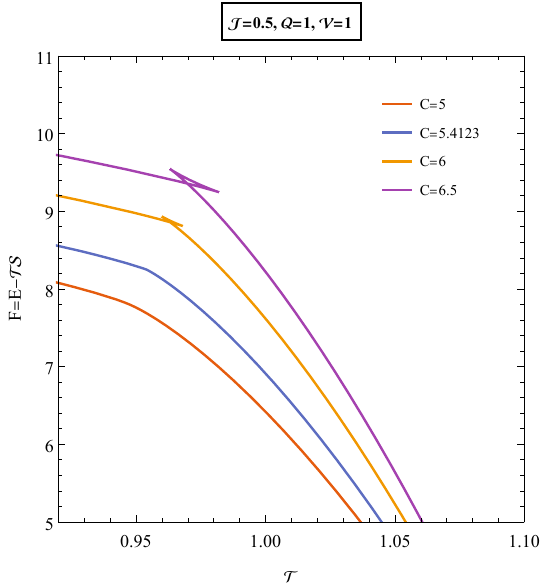}
\caption{}
\label{fig:plot3}
\end{subfigure}
    \caption{Free energy $F$ vs. temperature $\mathcal{T}$ plot for the fixed $(\mathcal{Q},\mathcal{J},\mathcal{V},C)$ ensemble in $(D=4/d=3)$.  }
\label{fig:en1}
\end{figure}

In this canonical ensemble, we look into fixing the electric charge $\mathcal{Q}$, the CFT volume $\mathcal{V}$, the angular momentum $\mathcal{J}$, and the central charge $C$. The Helmholtz free energy is given as:-
\begin{equation}
\label{eq:free1}
F=E-\mathcal{T}\mathcal{S}=\frac{16 \pi ^2 C^2 \left(12 \pi ^2 \mathcal{J}^2+\mathcal{S}^2\right)+16 \pi ^3 C \mathcal{S} \left(2 J^2+\mathcal{Q}^2\right)+2 \pi ^2 \mathcal{Q}^2 \mathcal{S}^2+3 \pi ^4 \mathcal{Q}^4-\mathcal{S}^4}{4 \pi  \sqrt{C \mathcal{S} \mathcal{V}} A_3}
\end{equation} 
and the temperature in this ensemble is given as:-
\begin{equation}
\mathcal{T}=
\frac{16 \pi ^2 C^2 \left(\mathcal{S}^2-4 \pi ^2 \mathcal{J}^2\right)+16 \pi  C \mathcal{S}^3+2 \pi ^2 \mathcal{Q}^2 \mathcal{S}^2-\pi ^4 \mathcal{Q}^4+3 \mathcal{S}^4}{4 \pi \mathcal{S}^{3/2} \sqrt{C \mathcal{V}}A_3}
\end{equation}
Where $A_3$ is given in Appendix \ref{App}.
From the CFT first law  \eqref{eq:first law 1}, the differentiation of $F$ satisfies-
\begin{equation}
\label{eq:diff1}
dF=dE-\mathcal{T} d\mathcal{S}-\mathcal{S}d\mathcal{T}=-\mathcal{S} d\mathcal{T}+\varphi d\mathcal{Q}-pd\mathcal{V}+\mu dC+\Omega d\mathcal{J}
\end{equation}
It can be see that $F$ is fixed at $(\mathcal{T}, \mathcal{Q},\mathcal{V},C,\mathcal{J})$.
We can now infer how the free energy $F$ behaves as a function of T for different fixed values of $(\mathcal{Q},\mathcal{V},C,\mathcal{J})$. For this, we can plot $F(T)$ parametrically using the entropy $\mathcal{S}$ as the parameter for the fixed values. In the figure \ref{fig:en1} we plot three different plots for different values of $\mathcal{Q}$, $\mathcal{J}$, $C$ while keeping the rest constant. We also plot the specific heat plots to check the stability of the entropy branches.   

\begin{figure}[htp]
\centering
\begin{subfigure}[b]{0.3\textwidth} 
\includegraphics[width=1\textwidth]{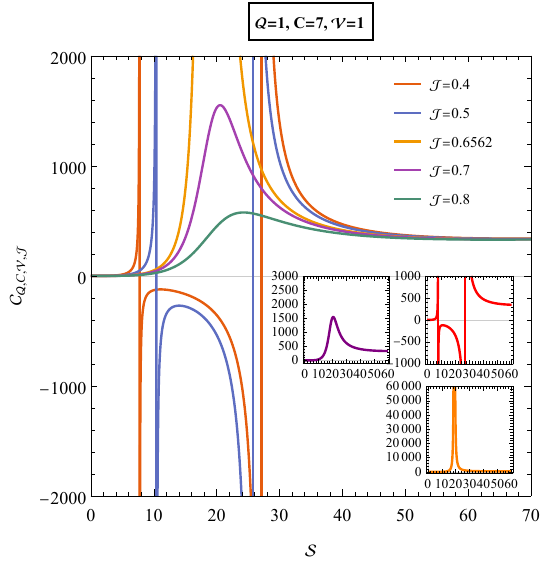}
\caption{}
\label{fig:plota}
\end{subfigure}
\begin{subfigure}[b]{0.3\textwidth}
\includegraphics[width=1\textwidth]{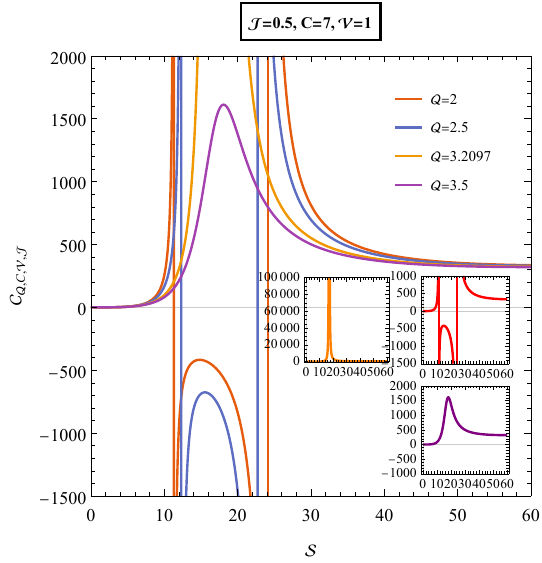}
\caption{}
\label{fig:plotb}
\end{subfigure}
\begin{subfigure}[b]{0.3\textwidth} 
\includegraphics[width=1\textwidth]{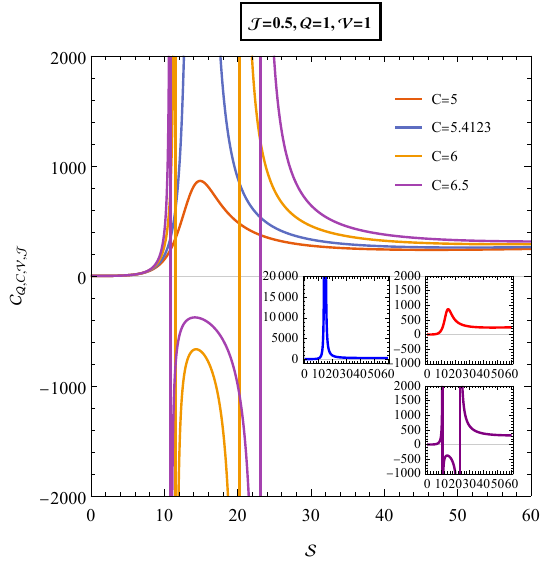}
\caption{}
\label{fig:plotc}
\end{subfigure}
    \caption{Specific heat $\mathcal{C}$ vs. entropy $\mathcal{S}$ plot for the fixed $(\mathcal{Q},\mathcal{J},\mathcal{V},C)$ ensemble in $(D=4/d=3)$.  }
\label{fig:ena}
\end{figure}
In Figure \ref{fig:plot1}, we study the $F$ vs $\mathcal{T}$ behavior for various values of $\mathcal{J}$. For subcritical values of $\mathcal{J}$ (red, blue curves), we observe a Van der Waals-like first-order phase transition, characterized by a swallow-tail structure. This transition exhibits three distinct black hole branches: small (stable), intermediate (unstable), and large (stable), as illustrated in Figure \ref{fig:plota}. At the critical value $\mathcal{J} = 0.6562$, a second-order phase transition emerges (orange curve), appearing as a kink in the $F$-$\mathcal{T}$ plot and a peak in the $\mathcal{C}$-$S$ plot (orange curve). This behavior is reminiscent of a superfluid $\lambda$-phase transition, analogous to phenomena observed in condensed matter systems, such as transitions from normal to superfluid states, superconductivity, and shifts from paramagnetism to ferromagnetism. The critical point in Figure \ref{fig:plotn} (orange curve) resembles the $\lambda$-shaped transition of Helium-4 from a normal fluid to a superfluid phase. This resemblance invites an interpretation of the small-to-large black hole transition as analogous to the transition from a normal fluid to a superfluid phase.This $\lambda$-like behavior persists in the $C$-$S$ plots across other ensembles. For supercritical values of $\mathcal{J}$ (purple curve), we observe a monotonous plot indicating a single, stable phase. In Figure \ref{fig:plot2}, we explore the system's behavior for various values of $\mathcal{Q}$. For subcritical values, we again find a first-order Van der Waals-type transition, with the characteristic three entropy branches (small-stable, intermediate-unstable, large-stable). At the critical value $\mathcal{Q} = 3.2097$, a superfluid $\lambda$-phase transition occurs (orange curve). For supercritical values of $\mathcal{Q}$, the system exhibits a monotonous single branch, representing a stable phase. Interestingly, when plotting for various values of $C$, the behavior reverses. Here, we observe a first-order Van der Waals-like transition with three branches for supercritical values of $C$, a $\lambda$-like  second-order phase transition at $C = 5.4123$, and a smooth, monotonous curve for subcritical values of $C$.
The stability of the different branches is determined through the specific heat, as shown in Figures \ref{fig:plota}, \ref{fig:plotb}, and \ref{fig:plotc}. A branch is stable if the specific heat is positive, while it becomes unstable when the specific heat is negative.

\begin{figure}[htp]
\centering
\begin{subfigure}[b]{0.3\textwidth} 
\includegraphics[width=1\textwidth]{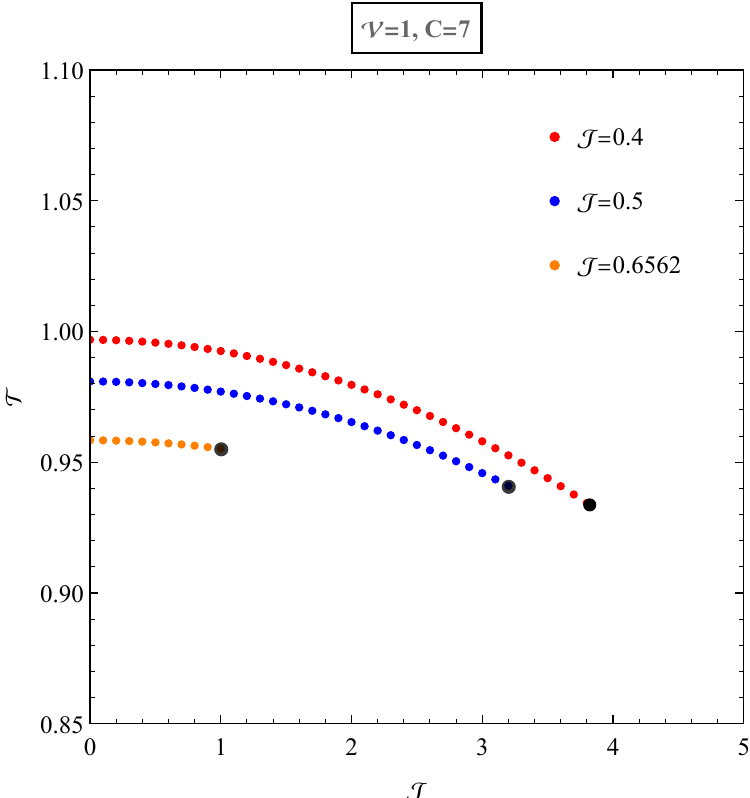}
\caption{}
\label{fig:plota1}
\end{subfigure}
\begin{subfigure}[b]{0.3\textwidth}
\includegraphics[width=1\textwidth]{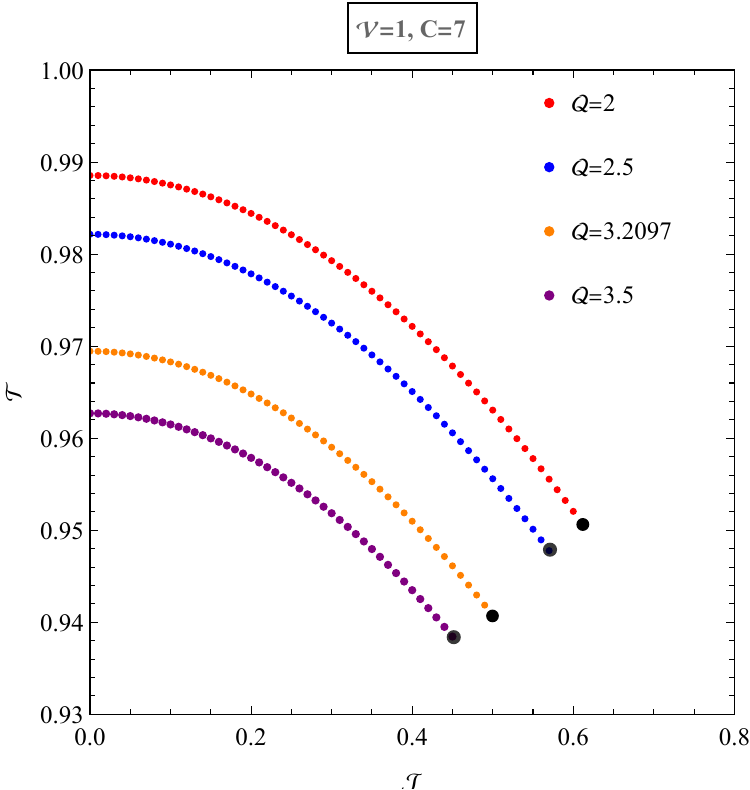}
\caption{}
\label{fig:plotb1}
\end{subfigure}
    \caption{$\bold{Left:}$Coexistence plots for various values of $\mathcal{J}$. $\bold{Right:}$ Coexistence plots for various values of $\mathcal{Q}$. The black dot represents the critical points in both the plots.}
\label{fig:ena1}
\end{figure}
The coexistence plots are seen in Figure \ref{fig:ena1}. Each plots is a line of phase transition of order one which stops at the critical points. The high entropy phases is in the right of the graphs while the low entropy phases is at the left side. After crossing the critical points (black dot) the two phases becomes indistinguishable.
\subsection{Ensemble at fixed $(\mathcal{Q},\Omega,\mathcal{V},C)$}
\label{sec:en2}
\begin{figure}[htbp]
\centering
\begin{subfigure}[b]{0.3\textwidth} 
\includegraphics[width=1\textwidth]{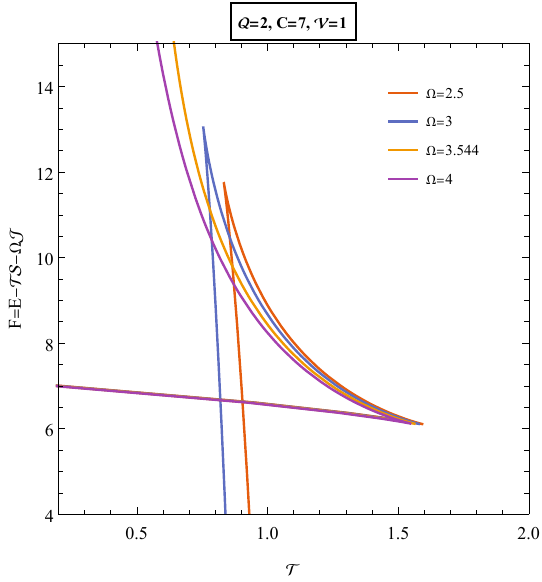}
\caption{}
\label{fig:plot4}
\end{subfigure}
\begin{subfigure}[b]{0.3\textwidth}
\includegraphics[width=1\textwidth]{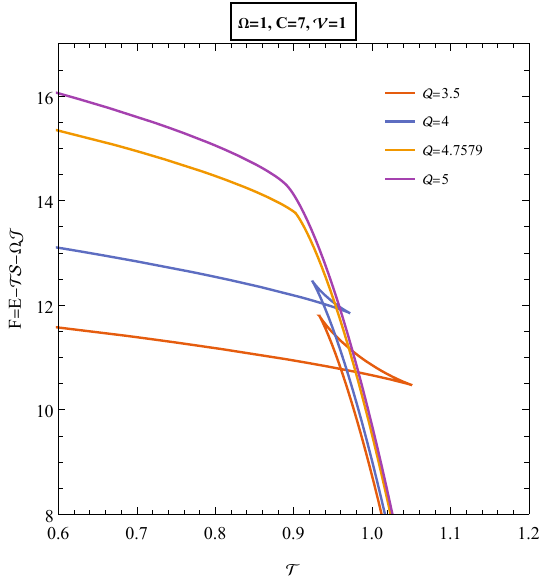}
\caption{}
\label{fig:plot5}
\end{subfigure}
\begin{subfigure}[b]{0.3\textwidth} 
\includegraphics[width=1\textwidth]{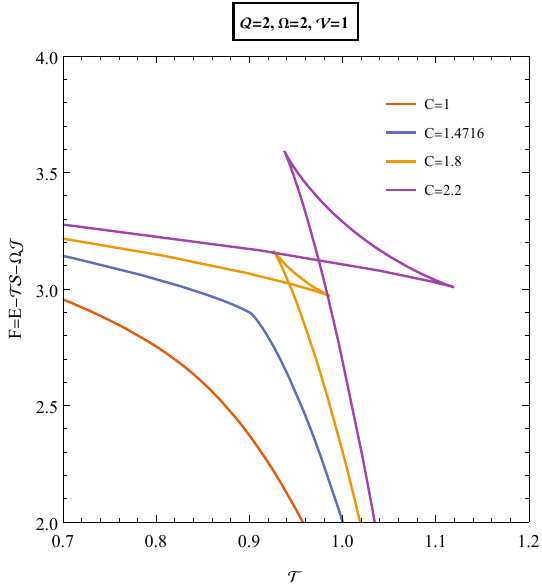}
\caption{}
\label{fig:plot6}
\end{subfigure}
    \caption{Free energy $F$ vs. temperature $\mathcal{T}$ plot for the fixed $(\mathcal{Q},\Omega,\mathcal{V},C)$ ensemble in $d=4$}
    \label{fig:en2}
\end{figure}

\begin{figure}[htp]
\centering
\begin{subfigure}[b]{0.3\textwidth} 
\includegraphics[width=1\textwidth]{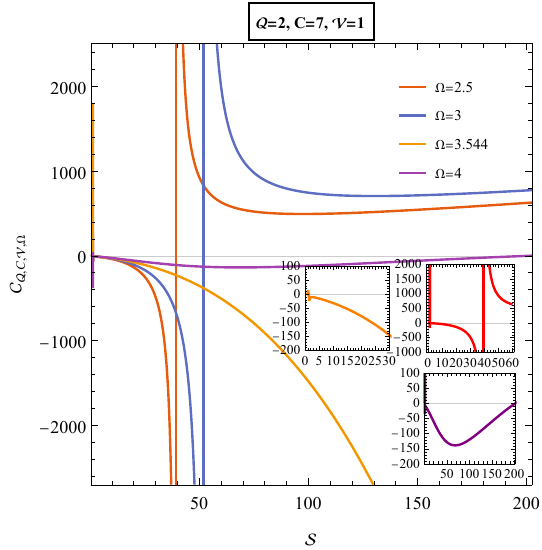}
\caption{}
\label{fig:plotd}
\end{subfigure}
\begin{subfigure}[b]{0.3\textwidth}
\includegraphics[width=1\textwidth]{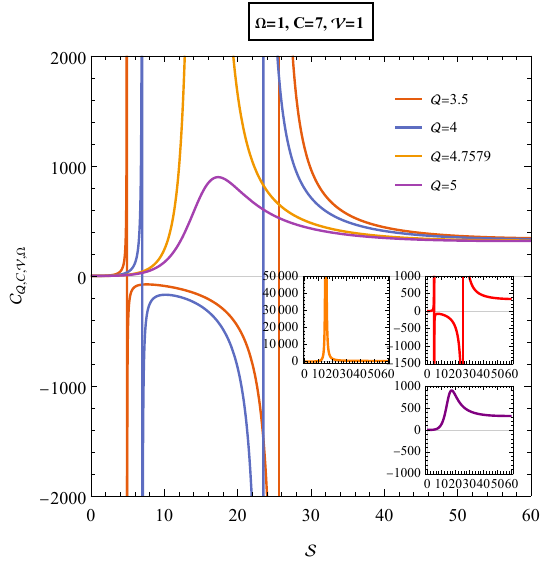}
\caption{}
\label{fig:plote}
\end{subfigure}
\begin{subfigure}[b]{0.3\textwidth} 
\includegraphics[width=1\textwidth]{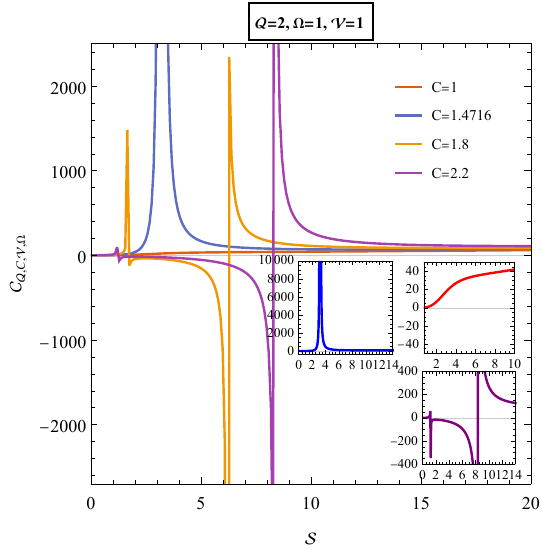}
\caption{}
\label{fig:plotf}
\end{subfigure}
    \caption{Specific heat $\mathcal{C}$ vs. entropy $\mathcal{S}$ plot for the fixed $(\mathcal{Q},\Omega,\mathcal{V},C)$ ensemble in $(D=4/d=3)$.  }
\label{fig:enb}
\end{figure}

In this grand canonical ensemble, we try to fix the angular velocity $\Omega$, the electric charge $\mathcal{Q}$, the CFT volume $\mathcal{V}$, and the central charge $C$. The free energy and temperature are:-

\begin{equation}
\label{eq:en2}
F=E-\mathcal{T} \mathcal{S}-\Omega \mathcal{J}=\frac{A_{35}-\frac{2  \sqrt{S} \sqrt{V} \Omega ^2 \left(4 \pi  C S+\pi ^2 Q^2+S^2\right)}{\sqrt{C} \sqrt{\left((4 \pi  C+S) \left(16 \pi ^2 C-S V \Omega ^2+4 \pi  S\right)\right)}}+8 \pi  \sqrt{\frac{(4 \pi  C+S) \left(4 \pi  C S+\pi ^2 Q^2+S^2\right)^2}{C S V \left(16 \pi ^2 C-S V \Omega ^2+4 \pi  S\right)}}}{8 \pi ^{3/2}}
\end{equation}
\begin{equation}
\mathcal{T}=\frac{A_3 \left(-32 \pi ^2 C^2 A_{36}+S^2 \left(3 S^2-\pi ^2 Q^2\right) \left(4 \pi -V \Omega ^2\right)\right)}{8 \pi ^{3/2} S (4 \pi  C+S)^2 \left(4 \pi  C S+\pi ^2 Q^2+S^2\right)}
\end{equation}
Where $A_3, A_{35},A_{36}$ are in Appendix \ref{App}. Comparing from the CFT first law \eqref{eq:first law 1}, and differentiating $F$ from \eqref{eq:en2} we get:-
\begin{equation}
\label{eq:diff2}
dF=dE-\mathcal{T}d\mathcal{S}-\mathcal{S}d \mathcal{T}-\Omega d\mathcal{J}-\mathcal{J}d\Omega=-\mathcal{S}d\mathcal{T}+\varphi d\mathcal{Q}-p d\mathcal{V}+\mu dC-\mathcal{J}d \Omega
\end{equation}
From the above equation \eqref{eq:diff2} we can see that $F$ is fixed at $(\mathcal{T},\mathcal{Q},\mathcal{V},C,\Omega)$
We can see that the free energy $F$ behaves as a function of the temperature $\mathcal{T}$ for the fixed values of $(\mathcal{Q},\Omega,\mathcal{V},C)$. For this ensemble, we can plot $F(\mathcal{T})$ parametrically by using the help of the entropy $\mathcal{S}$.
We plot for various values of $\Omega$ in Figure \ref{fig:plot4} and we see Van der Waals type phase transition of order one for subcritical values of $\Omega$ and see small (stable)-intermediate (unstable)-large (stable) entropy branches (red) in Figure \ref{fig:plotd}. At critical values $\Omega=3.544$ and onwards we see a Davies type phase transition where we see only two, one  small (stable) and one large (unstable) entropy branch (orange, purple) but no monotonous single branch. In Figure \ref{fig:plot5} we see for subcritical values of $\mathcal{Q}$ we see Van der Waals type phase transition showing 3 branches in Figure \ref{fig:plote} (red) and for $\mathcal{Q}=4.7579$ we obtain a second order superfluid $\lambda$- phase transition of order 2 and for supercritical values we get a monotonous single branch (purple). But in plot Figure \ref{fig:plot6} we see the transition in opposite order namely Van der Waals for supercritical values, second order $\lambda$ phase transition at the critical $C=1.4716$ and for subcritical points a monotonous single stable phase. Stability is seen in Figure \ref{fig:plotf}.

\subsection{Ensemble at fixed $(\varphi, \Omega,\mathcal{V},C)$}
\label{sec:en3}
In this ensemble we kind of fix the angular velocity $\Omega$, the electric potential $\Phi$, the CFT volume $\mathcal{V}$, and the central charge $C$. The free energy and temperature is labelled as:-
\begin{equation}
\label{eq:en313}
F=E-\mathcal{T} \mathcal{S}-\varphi \mathcal{Q}-\Omega \mathcal{J}=A_4
\end{equation}
and the temperature in this ensemble is given as:-
\begin{equation}
\label{eq:temp3}
\mathcal{T}=A_5
\end{equation}
where $A_4$ and $A_5$ are in Appendix \ref{App} as it is extremely long. If we see the CFT first law \eqref{eq:first law 1}, differentiating the free energy expression F \eqref{eq:en313}, we get:-
\begin{equation}
\label{eq:diff3}
dF=dE-\mathcal{T}d\mathcal{S}-\mathcal{S}d\mathcal{T}-\Omega d\mathcal{J}-\mathcal{J}d\Omega-\varphi d\mathcal{Q}-\mathcal{Q}d \varphi=-\mathcal{S}d\mathcal{T}-\mathcal{Q}d \varphi-p d\mathcal{V}+\mu dC-\mathcal{J}d\Omega
\end{equation}
From the above equation \eqref{eq:diff3} we can notice that the $F$ is kind of fixed at $(\mathcal{T},\varphi,\mathcal{V},C,\Omega)$. We see from the above equation \eqref{eq:diff3} that the free energy behaves as a function of temperature $\mathcal{T}$ for the fixed values of $(\varphi, \Omega,\mathcal{V}, C)$.
\begin{figure}[htbp]
\centering
\begin{subfigure}[b]{0.3\textwidth} 
\includegraphics[width=1\textwidth]{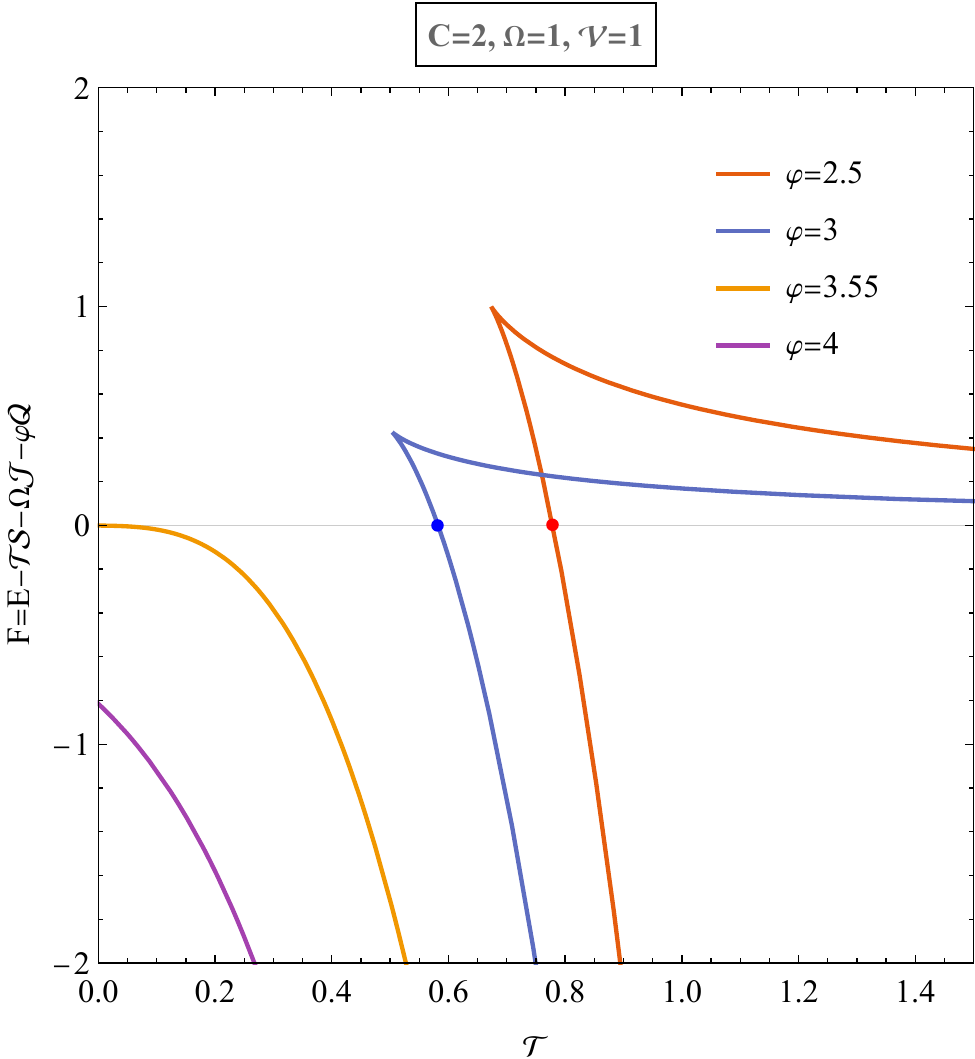}
\caption{}
\label{fig:plot7}
\end{subfigure}
\begin{subfigure}[b]{0.3\textwidth}
\includegraphics[width=1\textwidth]{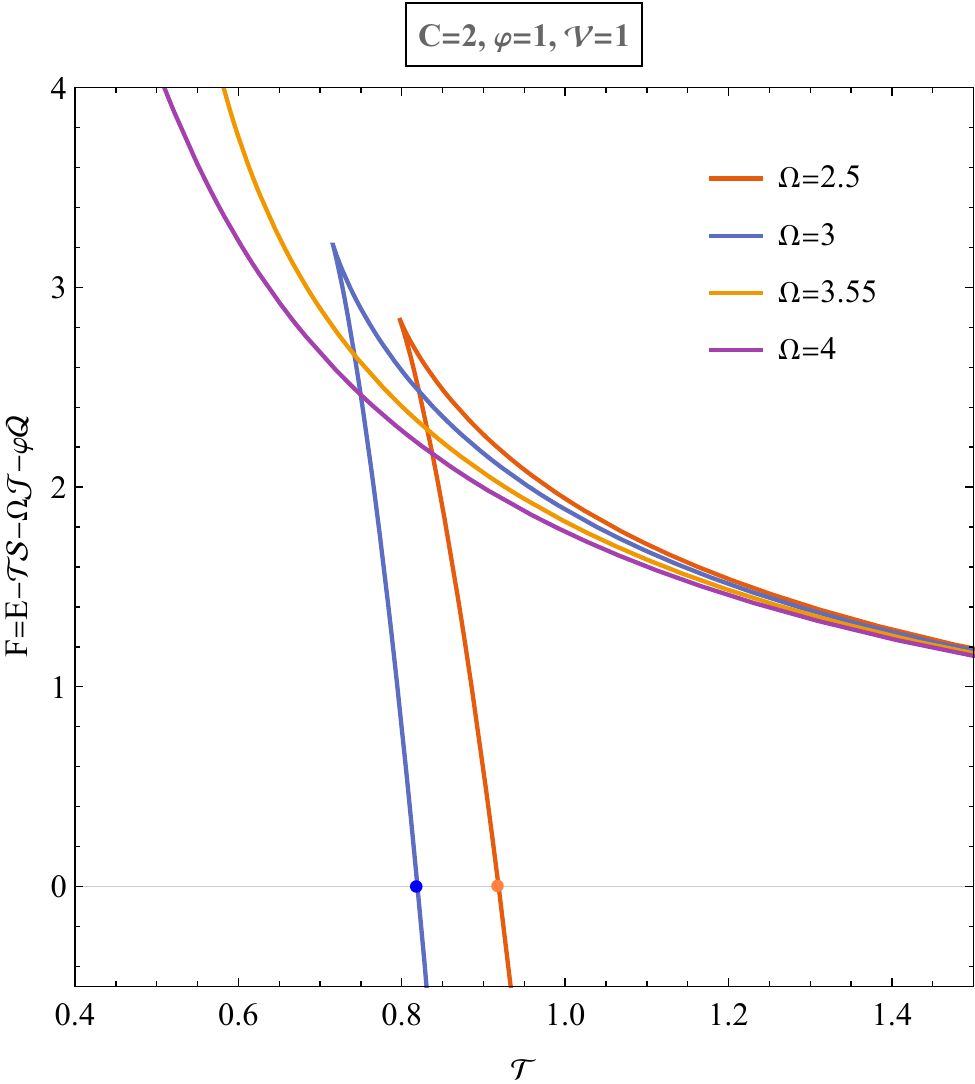}
\caption{}
\label{fig:plot8}
\end{subfigure}
\begin{subfigure}[b]{0.3\textwidth} 
\includegraphics[width=1\textwidth]{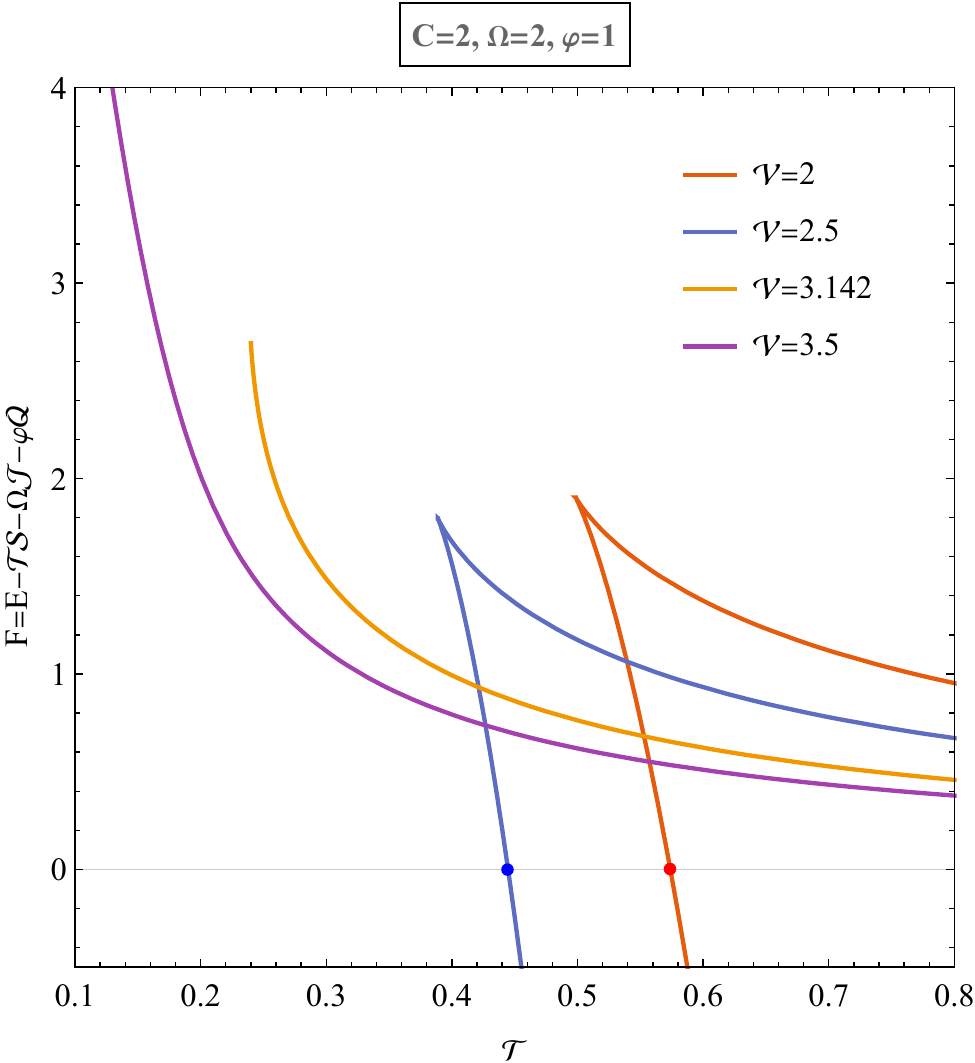}
\caption{}
\label{fig:plot9}
\end{subfigure}
    \caption{Free energy $F$ vs. temperature $\mathcal{T}$ plot for the fixed ($\varphi$,$\Omega$,$\mathcal{V}$,C) ensemble in $d=4$ dimensions}
    \label{fig:en3}
\end{figure}

\begin{figure}[htp]
\centering
\begin{subfigure}[b]{0.3\textwidth} 
\includegraphics[width=1\textwidth]{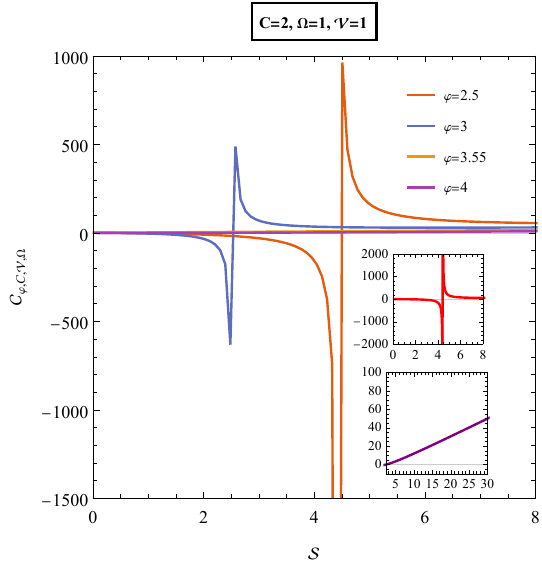}
\caption{}
\label{fig:plotg}
\end{subfigure}
\begin{subfigure}[b]{0.3\textwidth}
\includegraphics[width=1\textwidth]{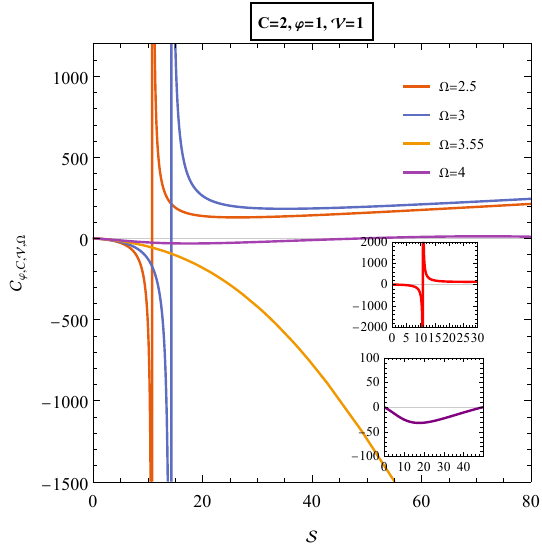}
\caption{}
\label{fig:ploth}
\end{subfigure}
\begin{subfigure}[b]{0.3\textwidth} 
\includegraphics[width=1\textwidth]{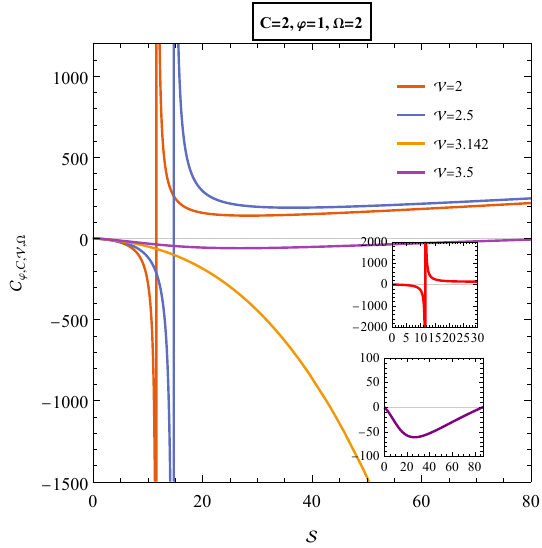}
\caption{}
\label{fig:ploti}
\end{subfigure}
    \caption{Specific heat $\mathcal{C}$ vs. entropy $\mathcal{S}$ plot for the fixed $(\varphi,\Omega,\mathcal{V},C)$ ensemble in $(D=4/d=3)$.  }
\label{fig:enc}
\end{figure}
\begin{figure}[htbp]
\centering
\includegraphics[scale=0.45]{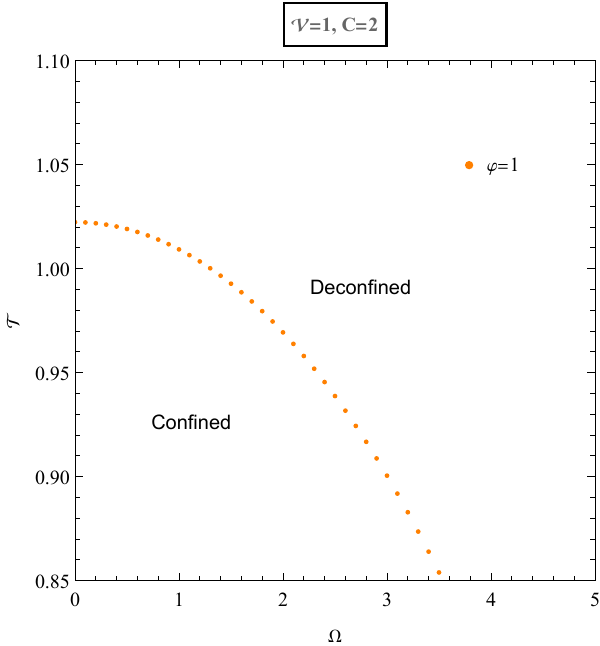}
\caption{$\Omega-\mathcal{T}$ phase diagram}
\label{fig:coexistp}
\end{figure}

From the expression of the free energy and temperature, we can plot the free energy $F(\mathcal{T})$ using the entropy $\mathcal{S}$ parametrically for the fixed value of $(\Omega,\varphi, \mathcal{V},C)$ as we can see in the Figure \ref{fig:en3}.
In figure \ref{fig:plot7} we plot for various values of $\varphi$ where for subcritical values of $\varphi$ (red, blue) we get two entropy branches of entropy namely small entropy (unstable) and large entropy (stable) branches which corresponds to small and large black holes. When the plot crosses the free energy (blue, red dot), it signals a first order phase transition named as (de)confinement phase transition which is analogous to Hawking-Page type phase transition. At critical value of $\varphi=3.55$ we see only one large (stable) entropy branch (orange). In figure \ref{fig:plotg} we see Davies type phase transition for $\varphi=2.5,3$ and rest we see a single stable phase. When the free energy $F<0$ then the large-entropy which is termed as "deconfined" seems to be dominating. But when $F>0$, we see that the "confined" state is more preferred. \\
In figure \ref{fig:plot8} we plot for various values of $\Omega$ where for subcritical values (red, blue) we see two branches namely small (unstable) and large (stable) entropy branches. Here also we get a (de)confinement phase transition and Davies type phase transition in figure \ref{fig:ploth}. Note that the blue and red dots are signalling (de)confinement phase transition and not Davies type. But at a critical value of $\Omega$ and so on we get only one (unstable) entropy branch. We can also calculate the turning point where the small (unstable) entropy analogous black hole transits to a large entropy analogous black hole. Considering the red plot and doing $\frac{\partial \mathcal{T}}{\partial \mathcal{S}}=0$ we get the turning point entropy at $\mathcal{S}_{cusp}=10.8421$ and then putting it in $\mathcal{T}$, we get $\mathcal{T}_{cusp}=0.743092$ in figure \ref{fig:plot8}, \ref{fig:ploth}. This point is actually where the Davies phase transition occurs and is known as the Davies point.
In figure \ref{fig:plot9}, we see for $\mathcal{V}$=2, 2.5 we see a (de)confinement phase transition and the transition points are marked in blue and red dots. With the same values in figure \ref{fig:ploti}, we see Davies type phase transition. After crossing the critical value $\mathcal{V}=3.142$ we see only an unstable branch with no phase transition. In figure \ref{fig:coexistp} we see the $\Omega-\mathcal{T}$ phase diagram with the coexistence curve which represents the line of (de)confinement phase transitions in the CFT. 

\subsection{Ensemble at fixed $(\varphi,\mathcal{J},\mathcal{V},C)$}
\label{sec:en4}
In this ensemble we try to fix the electric potential $\varphi$, the angular velocity $\mathcal{J}$, CFT volume $\mathcal{V}$, and the central charge $C$. The free energy is given as:-

\begin{equation}
\label{en:eq4}
\begin{split}
&F=E-\mathcal{T}\mathcal{S}-\varphi \mathcal{Q}=\frac{\sqrt{\frac{64 \pi ^4 C^2 \mathcal{J}^2+16 \pi ^2 C^2 \mathcal{S}^2+\frac{8 \pi  C D_2 \mathcal{S}}{3 D_1}+16 \pi ^3 C \mathcal{J}^2 \mathcal{S}+8 \pi  C \mathcal{S}^3+\frac{2 D_2 \mathcal{S}^2}{3 D_1}+\frac{D_2^2}{D_3}+\mathcal{S}^4}{C \mathcal{S} \mathcal{V}}}}{2 \pi }\\
&-\frac{16 \pi ^2 C^2 \left(\mathcal{S}^2-4 \pi ^2 \mathcal{J}^2\right)+16 \pi  C \mathcal{S}^3+\frac{2 D_2 \mathcal{S}^2}{3 D_1}-\frac{D_2^2}{D_3}+3 \mathcal{S}^4}{4 \pi  C \mathcal{S} \mathcal{V} \sqrt{\frac{16 \pi ^2 C^2 \left(4 \pi ^2 \mathcal{J}^2+\mathcal{S}^2\right)+8 \pi  C \mathcal{S} \left(\frac{D_2}{3 D_1}+2 \pi ^2 \mathcal{J}^2+\mathcal{S}^2\right)+\left(\frac{D_2}{3 D_1}+\mathcal{S}^2\right){}^2}{C \mathcal{S} \mathcal{V}}}}-\frac{\sqrt{\frac{D_2}{D_1}} \varphi }{\sqrt{3} \pi }
\end{split}
\end{equation} and the temperature is:-
\begin{equation}
\mathcal{T}=\frac{16 \pi ^2 C^2 \left(\mathcal{S}^2-4 \pi ^2 \mathcal{J}^2\right)+16 \pi  C \mathcal{S}^3+\frac{2 D_2 \mathcal{S}^2}{3 D_1}-\frac{D_2^2}{D_3}+3 \mathcal{S}^4}{4 \pi  C \mathcal{S}^2 \mathcal{V} \sqrt{\frac{16 \pi ^2 C^2 \left(4 \pi ^2 \mathcal{J}^2+\mathcal{S}^2\right)+8 \pi  C \mathcal{S} \left(\frac{D_2}{3 D_1}+2 \pi ^2 \mathcal{J}^2+\mathcal{S}^2\right)+\left(\frac{D_2}{3 D_1}+\mathcal{S}^2\right){}^2}{C \mathcal{S} \mathcal{V}}}}
\end{equation}
Where $D_1$, $D_2$, $D_3$ are given in Appendix \ref{App}.
From the CFT first law, we notice that  from the free energy expression \eqref{eq:first law 1}, we put the value in the result when differentiating \eqref{en:eq4}, we get:-
\begin{equation}
\label{eq:diff4}
dF=dE-\mathcal{T}d\mathcal{S}-\mathcal{S}d\mathcal{T}-\varphi d\mathcal{Q}-\mathcal{Q}d \varphi=-\mathcal{S}d\mathcal{T}-\mathcal{Q}d\varphi-p d\mathcal{V}+\mu dC
\end{equation}
We see that the free energy $F$ \eqref{eq:diff4} is a kind of fixed at $(\mathcal{T},\varphi,\mathcal{V},C,\Omega)$. We can note that the free energy $F$ behaves as a function of the temperature $\mathcal{T}$ for the various fixed values of $(\varphi,\mathcal{J},\mathcal{V},C)$. For this system, we can plot parametrically $F(\mathcal{T})$ by using the entropy $\mathcal{S}$ and the specific heat plot.  
\begin{figure}[htbp]
\centering
\begin{subfigure}[b]{0.3\textwidth} 
\includegraphics[width=1\textwidth]{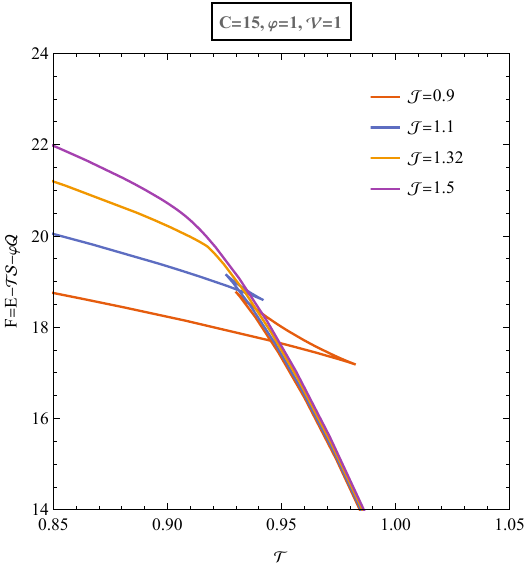}
\caption{}
\label{fig:plot10}
\end{subfigure}
\begin{subfigure}[b]{0.3\textwidth}
\includegraphics[width=1\textwidth]{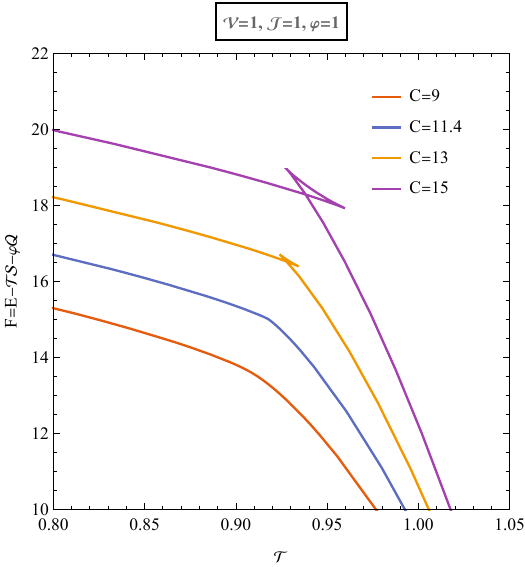}
\caption{}
\label{fig:plot11}
\end{subfigure}
\begin{subfigure}[b]{0.3\textwidth}
\includegraphics[width=1\textwidth]{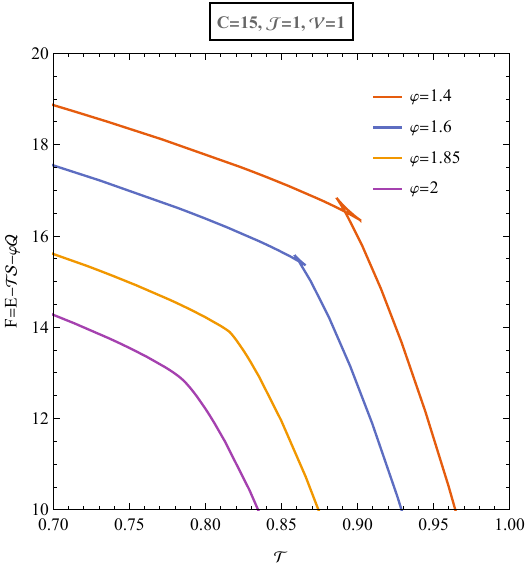}
\caption{}
\label{fig:plot11b}
\end{subfigure}
    \caption{Free energy F vs temperature $\mathcal{T}$ plot for fixed ($\varphi$,$\mathcal{J}$,$\mathcal{V}$,C) ensemble in $d=4$}
    \label{fig:en4}
\end{figure}
\begin{figure}[htp]
\centering
\begin{subfigure}[b]{0.3\textwidth} 
\includegraphics[width=1\textwidth]{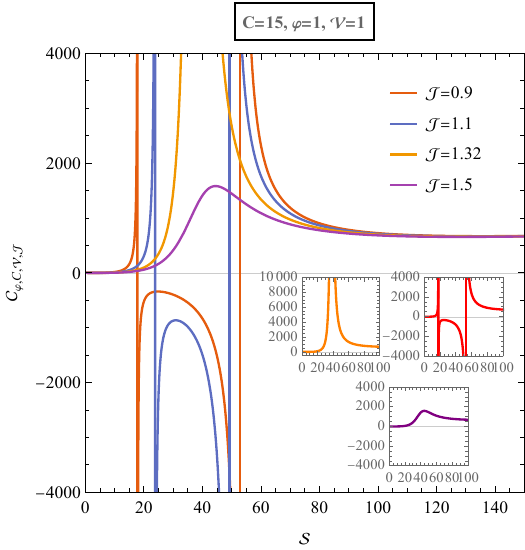}
\caption{}
\label{fig:plotj}
\end{subfigure}
\begin{subfigure}[b]{0.3\textwidth}
\includegraphics[width=1\textwidth]{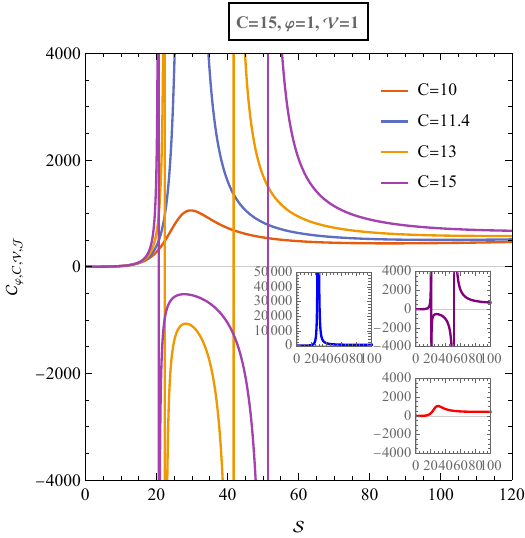}
\caption{}
\label{fig:plotk}
\end{subfigure}
\begin{subfigure}[b]{0.3\textwidth} 
\includegraphics[width=1\textwidth]{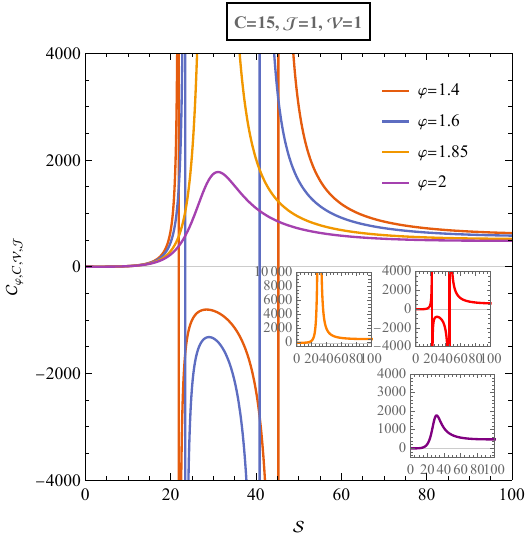}
\caption{}
\label{fig:plotl}
\end{subfigure}
    \caption{Specific heat $\mathcal{C}$ vs. entropy $\mathcal{S}$ plot for the fixed $(\varphi,\mathcal{J},\mathcal{V},C)$ ensemble in $(D=4/d=3)$.  }
\label{fig:end}
\end{figure}
In Figure \ref{fig:plot10}, we plot for different values of $\mathcal{J}$ where for subcritical values of $\mathcal{J}$ we get Van der Waals type phase transition of order one where we get small (stable)-intermediate (unstable)-large (stable) entropy branches also in figure \ref{fig:plotj} (red, blue). At critical value of $\mathcal{J}=1.32$ we get a a superfluid $\lambda$ phase transition of second order (orange) and for supercritical values we get a single phase of stable branch (purple). At figure \ref{fig:plot11b}, we plot for $\varphi$ where we see Van der Waals transition for subcritical values (red, blue) having small (stable)-intermediate (unstable)-large (stable) entropy branches from left to right , second order superfluid $\lambda$ phase transition (orange) at $\varphi=1.85$ and for supercritical values we get stable branch (purple) in figure \ref{fig:plotl}. But its the opposite in figure \ref{fig:plot11} where for supercritical values we get Van der Waals phase transition (purple, orange), then at $C=11.4$ we get a second order superfluid $\lambda$ phase transition (blue) and for subctitical values we get a single stable entropy branch (red) seen in figure \ref{fig:plotk}. 
\subsection{Ensembles at fixed $(\mathcal{Q},\Omega,p,C)$}
\label{sec:p1}
\begin{figure}[htbp]
\centering
\begin{subfigure}[b]{0.3\textwidth} 
\includegraphics[width=1\textwidth]{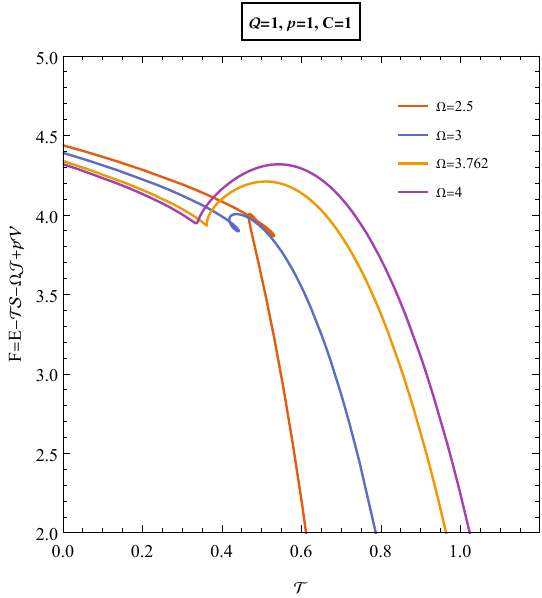}
\caption{}
\label{fig:plotQP}
\end{subfigure}
\begin{subfigure}[b]{0.3\textwidth}
\includegraphics[width=1\textwidth]{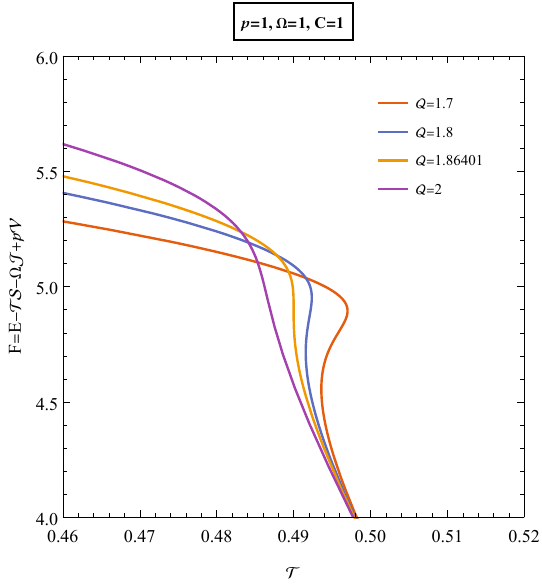}
\caption{}
\label{fig:plotQP1}
\end{subfigure}
\begin{subfigure}[b]{0.3\textwidth}
\includegraphics[width=1\textwidth]{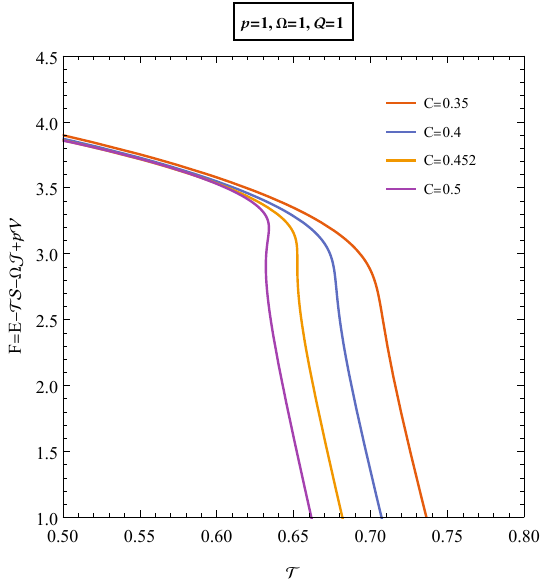}
\caption{}
\label{fig:plotQP2}
\end{subfigure}
    \caption{Free energy F vs temperature $\mathcal{T}$ plot for fixed ($\mathcal{Q}$,$\Omega$,$p$,C) ensemble in $d=4$}
    \label{fig:en5}
\end{figure}

\begin{figure}[htp]
\centering
\begin{subfigure}[b]{0.3\textwidth} 
\includegraphics[width=1\textwidth]{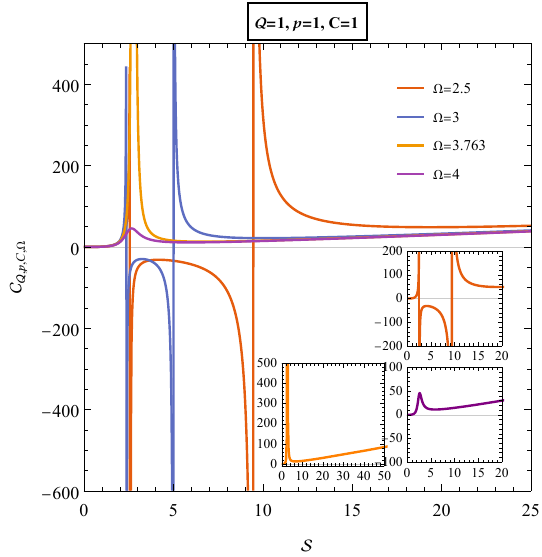}
\caption{}
\label{fig:plotm}
\end{subfigure}
\begin{subfigure}[b]{0.3\textwidth}
\includegraphics[width=1\textwidth]{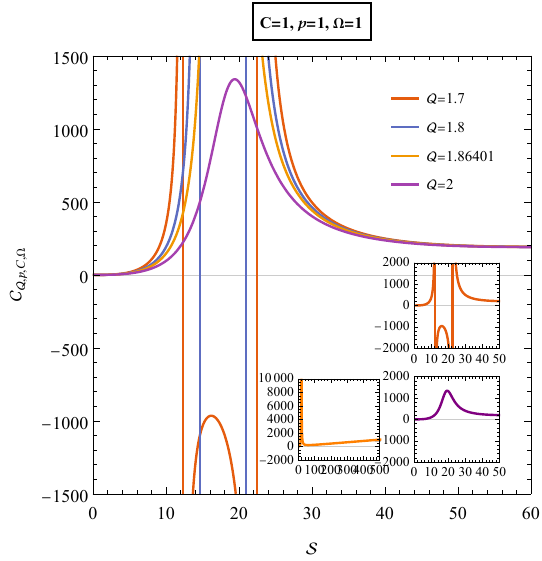}
\caption{}
\label{fig:plotn}
\end{subfigure}
\begin{subfigure}[b]{0.3\textwidth} 
\includegraphics[width=1\textwidth]{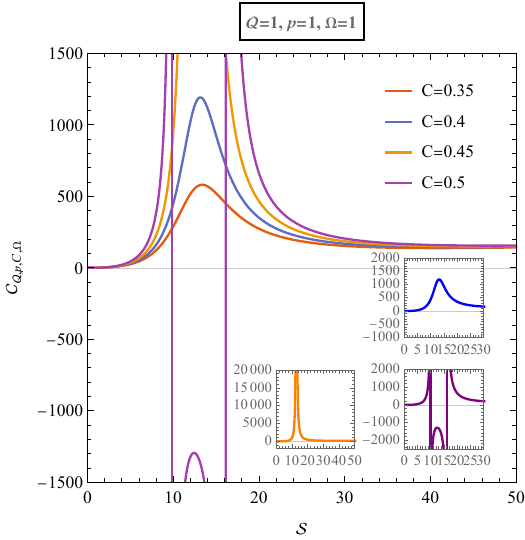}
\caption{}
\label{fig:ploto}
\end{subfigure}
  \caption{Specific heat $\mathcal{C}$ vs. entropy $\mathcal{S}$ plot for the fixed $(\mathcal{Q},\Omega, p,C)$ ensemble in $(D=4/d=3)$.}
\label{fig:ene}
\end{figure}

\begin{figure}[htp]
\centering
\begin{subfigure}[b]{0.3\textwidth} 
\includegraphics[width=1\textwidth]{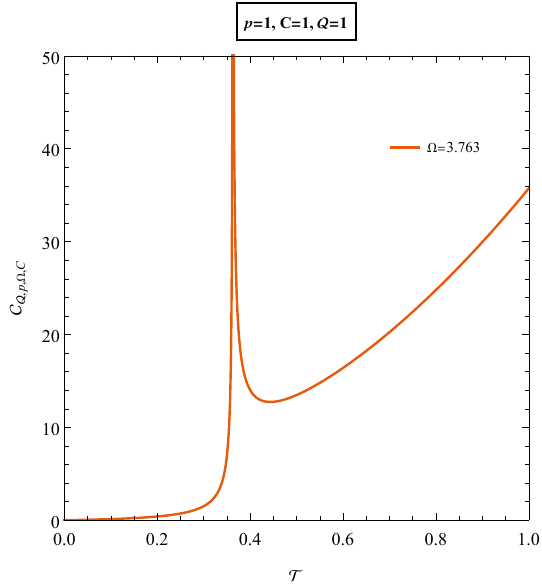}
\caption{}
\label{fig:CT1}
\end{subfigure}
\begin{subfigure}[b]{0.3\textwidth}
\includegraphics[width=1\textwidth]{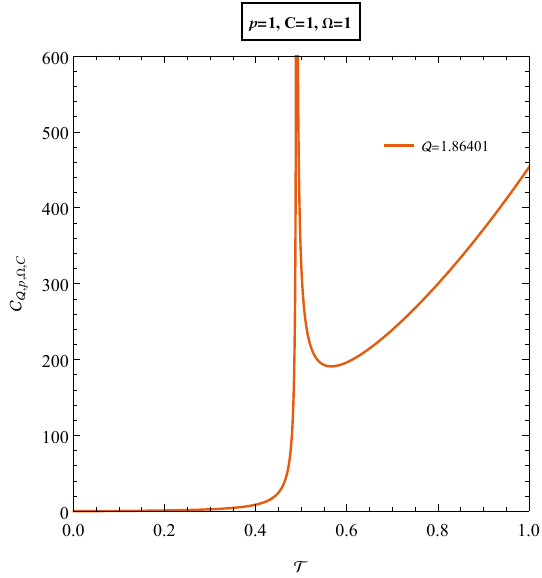}
\caption{}
\label{fig:CT2}
\end{subfigure}
\begin{subfigure}[b]{0.3\textwidth} 
\includegraphics[width=1\textwidth]{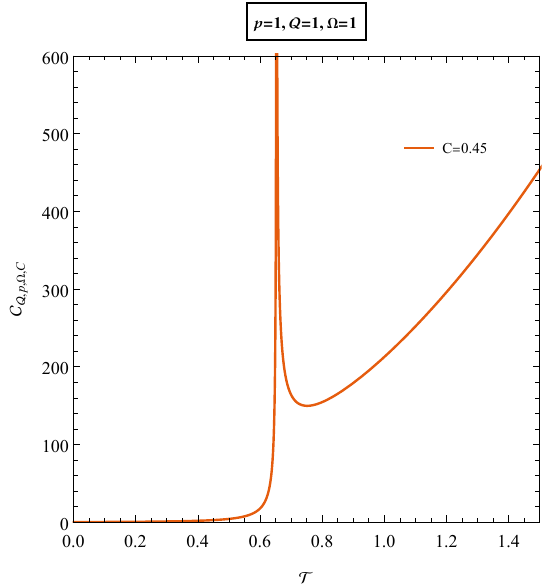}
\caption{}\label{fig:CT3}
\end{subfigure}
    \caption{Specific heat vs temperature for various critical values of $\mathcal{Q},\Omega,C$  where temperature diverges at the critical point showing a $\lambda$ like structure.}
\label{fig:CT}
\end{figure}

As seen in other CFT thermodynamic papers, there was no criticality found in the $p$ containing ensembles but here its is an execption as we see criticality in the variables taken into consideration. This must be due to the dual interplay of both electric charge and angular momentum.
In this ensemble we write the free energy as:-
\begin{equation}
\begin{split}
&F=E-\mathcal{T}\mathcal{S}-\Omega \mathcal{J}+p\mathcal{V}=\sqrt{\frac{2}{\pi }} E_2
+\frac{p \left(16 \pi ^2 C+4 \pi  \mathcal{S}-E_1\right)}{2 \mathcal{S} \Omega ^2}\\
&-\frac{ \sqrt{\mathcal{S}} \left(\pi ^2 \mathcal{Q}^2+\mathcal{S}^2+4 C \pi  \mathcal{S}\right) \Omega ^2 \sqrt{\frac{16 \pi ^2 c+4 \pi  \mathcal{S}-E_1}{\mathcal{S} \Omega ^2}}}{4 \sqrt{2} \sqrt{C} \pi ^{3/2} \sqrt{\left((4 \pi  C+\mathcal{S}) \left(16 \pi ^2 C+4 \pi  \mathcal{S}+\frac{1}{2} \left(-16 \pi ^2 C-4 \pi  \mathcal{S}+E_1\right)\right)\right)}}\\
&-\frac{\Omega ^2\left(16 \pi ^2 C^2 \left(\mathcal{S}^2-\frac{\left(4 \pi  C \mathcal{S}+\pi ^2 \mathcal{Q}^2+\mathcal{S}^2\right)^2\left(16 \pi ^2 C+4 \pi  \mathcal{S}-E_1\right)}{8 \pi  C (4 \pi  C+\mathcal{S}) \left(\frac{1}{2} \left(-16 \pi ^2 C-4 \pi  \mathcal{S}+E_1\right)+16 \pi ^2 C+4 \pi  \mathcal{S}\right)}\right)+16 \pi  C \mathcal{S}^3+2 \pi ^2 \mathcal{Q}^2 \mathcal{S}^2-\pi ^4 \mathcal{Q}^4+3 \mathcal{S}^4\right)}{4 \sqrt{2} \pi ^{3/2} C \left(16 \pi ^2 C+4 \pi  \mathcal{S}-E_1\right) E_2}
\end{split}
\end{equation}
and the temperature is written as:-
\begin{equation}
\mathcal{T}=\frac{E_2 \left(256 \pi ^4 \mathcal{S} C^3-32 \pi ^2 \left(2 \pi ^3 \mathcal{Q}^2-10 \pi  \mathcal{S}^2+\frac{1}{2} \mathcal{S} \left(16 \pi ^2 C+4 \pi  \mathcal{S}-E_1\right)\right) C^2-4 \pi  \mathcal{S} E_3 C+E_4\right)}{4 \sqrt{2} \pi ^{3/2} \mathcal{S} (4 \pi  C+\mathcal{S})^2 \left(\pi ^2 \mathcal{Q}^2+\mathcal{S}^2+4 C \pi  \mathcal{S}\right)}
\end{equation}
Where $E_1$, $E_2$, $E_3$, $E_4$ are given in Appendix \ref{App}.
Using the CFT first law and differentiating the free energy exression given above we get:-
\begin{equation}
dF=dE-\mathcal{T}d\mathcal{S}-\mathcal{S}d\mathcal{T}-\Omega d\mathcal{J}-\mathcal{J}d\Omega +p d\mathcal{V}+\mathcal{V}dp=-\mathcal{S}d \mathcal{T}-\varphi d\mathcal{Q}-\mathcal{J}d \Omega+\mathcal{V}dp-\mu dC
\end{equation}
We can note that the free energy $F$ behaves as a function of $\mathcal{T}$ for the various values of $(\mathcal{Q},\Omega,p,C)$. We plot the parametric plot $F(\mathcal{T})$ w.r.t to $\mathcal{S}$ in figure \ref{fig:en5} and specific heat plots in figure \ref{fig:ene}. We plot for various values of $\Omega$ in figure \ref{fig:plotQP} were for  subcritical values of $\Omega$ we get Van der Waals type phase transition of order one. Here in figure \ref{fig:plotm} we see three branches namely small (stable)-intermediate (unstable)-large (stable) entropy branches clearly (red, blue). At the critical point at $\Omega=3.762$ we get a second order superfluid $\lambda$ phase transition and we see two entropy branches namely small (stable)-large(stable) branches respectively in figure \ref{fig:plotm}. For supercritical values we get only one stable branch (purple). In figure \ref{fig:plotQP1}, \ref{fig:plotn} we plot for various values of electric charge $\mathcal{Q}$ where for subcritical values (red, blue) we get Van der Waals type phase transition of order one and at the critical value $\mathcal{Q}=1.8604$ we get the superfluid $\lambda$ phase transition of order two (orange) and for $\mathcal{Q}=2$ we get a single stable phase. Similarly for fig. \ref{fig:plotQP2}, \ref{fig:ploto} we plot for various values of central charge $C$ where for the critical values of $C=0.45$ we get the superfluid $\lambda$ phase transition. But for supercritical values of $C$ we get the first order Van der Waals phase transition and for subcritical values we see a single stable line (blue) . We can see the $\lambda$ like structure clearly for all critical values at figure \ref{fig:CT1}, \ref{fig:CT2}, \ref{fig:CT3}.

\subsection{Ensemble at fixed $(\varphi,\Omega,p,C)$}
\label{sec:p2}
In this ensemble the free energy is given as:-
\begin{equation}
\begin{split}
&F=E-\mathcal{T}\mathcal{S}-\varphi \mathcal{Q}+p\mathcal{V}=\frac{1}{8} \left(-8 F_1 \varphi +16 \pi ^{3/2} \sqrt{2} F_2\right.\\
&\left.-\frac{2 \left(-\mathcal{S} \left(\mathcal{S}^2+4 C \pi  \mathcal{S}+\pi ^2 F_1^2\right) \varphi ^3+8 p \pi ^3 (4 \pi  C+\mathcal{S})F_1+F_4\right)}{\pi ^2 \mathcal{S} \Phi ^2 F_1}\right.\\
&\left.-\frac{2  \sqrt{\mathcal{S}} \varphi  \Omega  \left(\mathcal{S}^2+4 C \pi  \mathcal{S}+\pi ^2 F_1^2\right) \sqrt{\frac{-\mathcal{S} \left(\mathcal{S}^2+4 C \pi  \mathcal{S}+\pi ^2 F_1^2\right) \varphi ^3+8 p \pi ^3 (4 \pi  C+\mathcal{S}) F_1+F_4}{p \mathcal{S} \varphi ^2 F_1}}}{\sqrt{C} \pi ^{3/2} \sqrt{\frac{(4 \pi  C+\mathcal{S}) F_3}{p F_1}}}\right.\\
&\left.-\frac{\sqrt{2} p \varphi ^2 F_1 \left(3 \mathcal{S}^4+16 C \pi  \mathcal{S}^3+16 C^2 \pi ^2 \mathcal{S}^2+2 \pi ^2 F_1^2 \mathcal{S}^2-\pi ^4 F_1^4-F_5\right)}{C \pi ^{3/2} F_2 \left(-\mathcal{S} \left(\mathcal{S}^2+4 C \pi  \mathcal{S}+\pi ^2 F_1^2\right) \varphi ^3+8 p \pi ^3 (4 \pi  C+\mathcal{S}) F_1+F_4\right)}\right)
\end{split}
\end{equation}

and the temperature is given as:-
\begin{equation}
\mathcal{T}=\frac{F_2 F_6}{8 \sqrt{2} p \pi ^{3/2} \mathcal{S} (4 \pi  C+\mathcal{S})^2 F_1\left(\mathcal{S}^2+4 C \pi  \mathcal{S}+\pi ^2 F_1^2\right)}
\end{equation}
Where $F_1$, $F_2$, $F_3$, $F_4$, $F_5$, $F_6$ are in Appendix \ref{App}.
Using the CFT first law and differentiating the free energy expression given above we get:-
\begin{equation}
dF=dE-\mathcal{T}d \mathcal{S}-\mathcal{S}d\mathcal{T}-\varphi d\mathcal{Q}-\mathcal{Q}d \varphi+p d\mathcal{V}+\mathcal{V}dp=-\mathcal{S}d\mathcal{T}-\mathcal{Q}d \varphi-\Omega d\mathcal{J}+\mathcal{V}dp-\mu dC
\end{equation}
\begin{figure}[htp]
\centering
\begin{subfigure}[b]{0.3\textwidth} 
\includegraphics[width=1\textwidth]{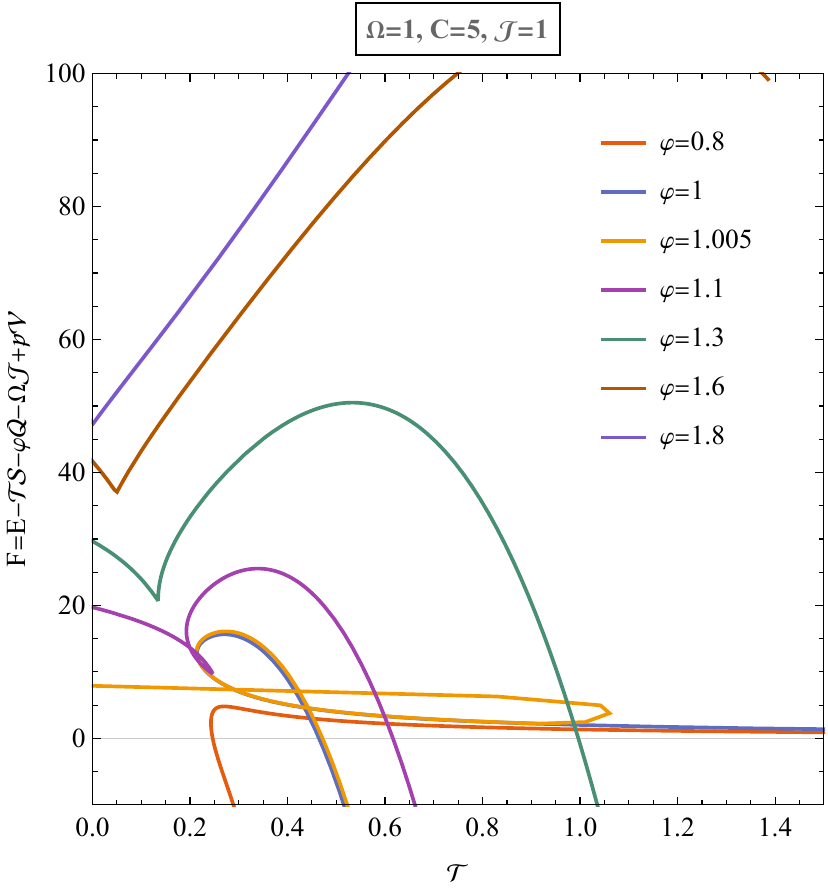}
\caption{}
\label{fig:plotomegap}
\end{subfigure}
\begin{subfigure}[b]{0.3\textwidth}
\includegraphics[width=1\textwidth]{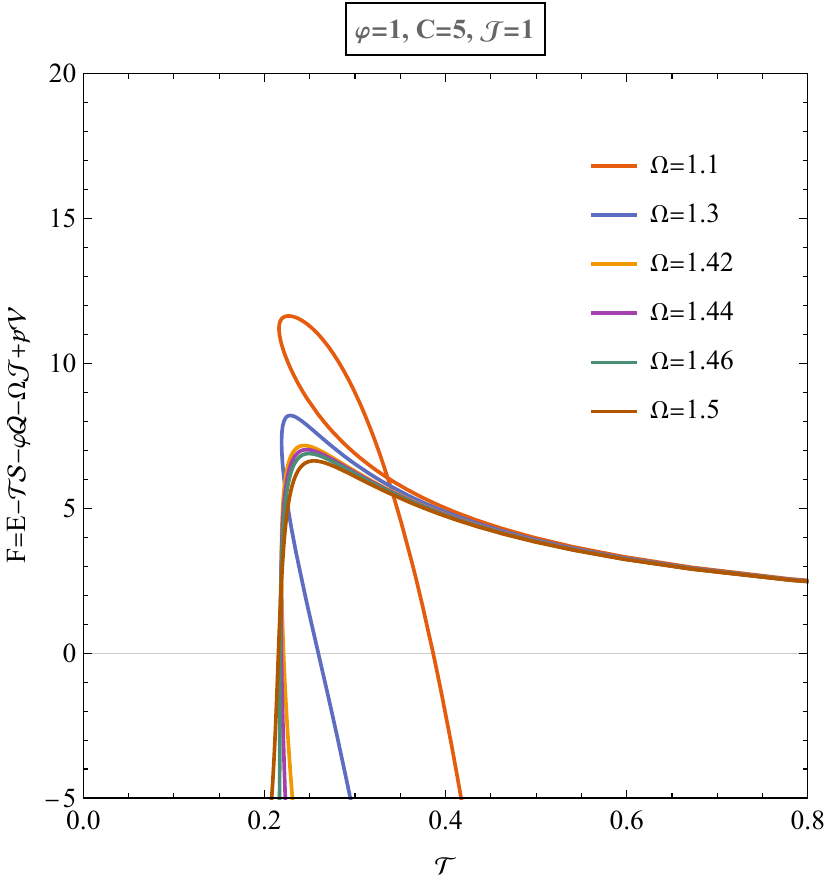}
\caption{}
\label{fig:plotomegap1}
\end{subfigure}
    \caption{Free energy F vs temperature $\mathcal{T}$ plot for fixed ($\varphi$, $\Omega$, $p$, $C$) ensemble in $D=4/d=3$}
    \label{fig:en6}
\end{figure}
\begin{figure}[htp]
\centering
\begin{subfigure}[b]{0.3\textwidth} 
\includegraphics[width=1\textwidth]{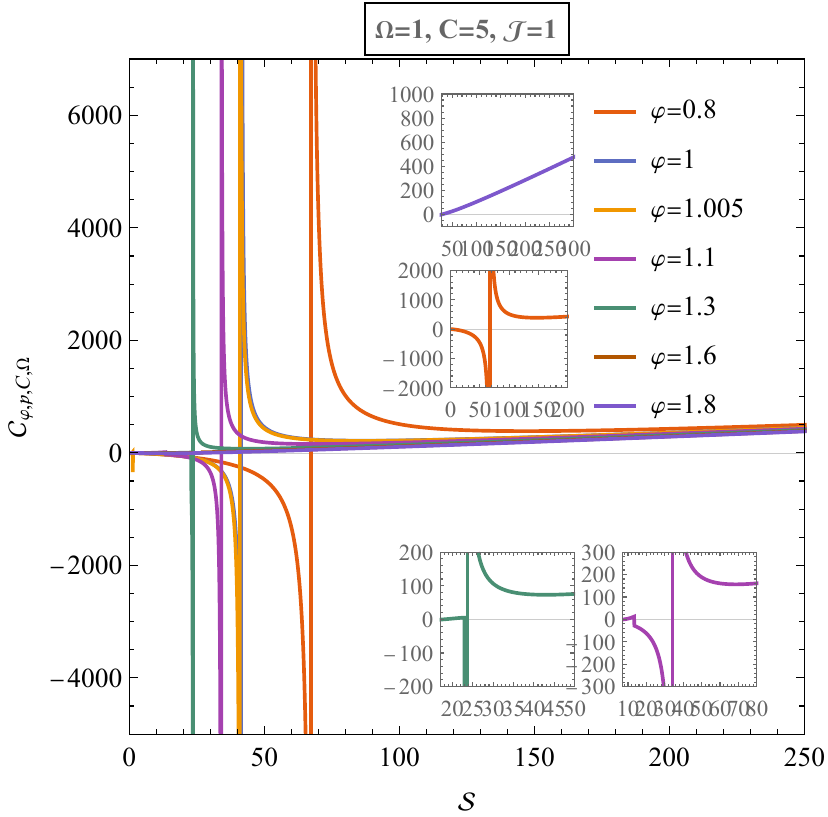}
\caption{}
\label{fig:plotp}
\end{subfigure}
\begin{subfigure}[b]{0.3\textwidth}
\includegraphics[width=1\textwidth]{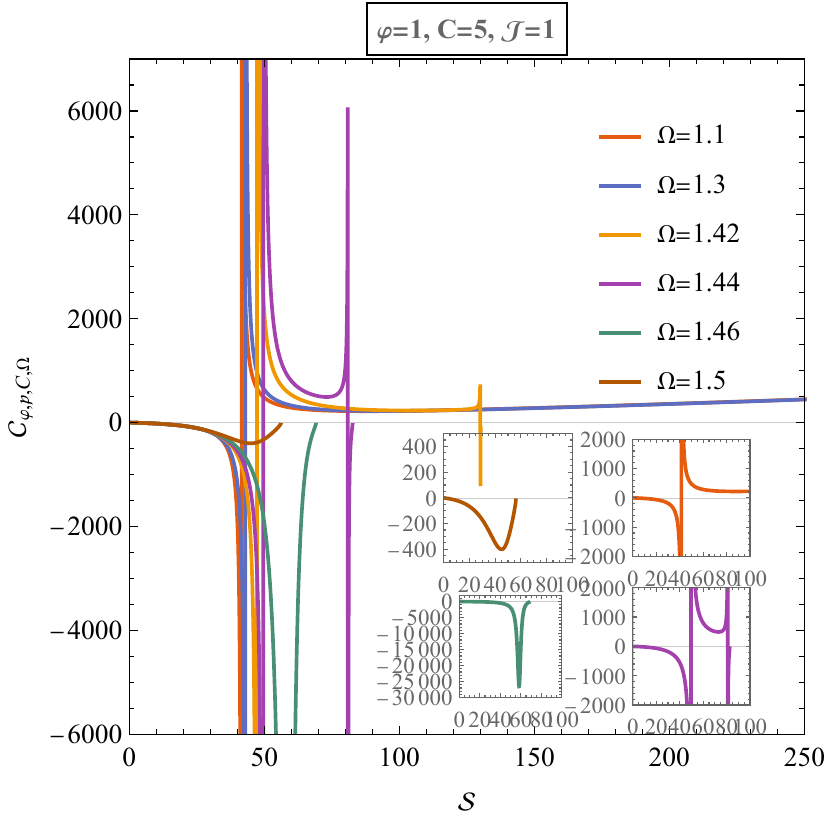}
\caption{}
\label{fig:plotq}
\end{subfigure}
    \caption{Specific heat $\mathcal{C}$ vs. entropy $\mathcal{S}$ plot for the fixed $(\varphi,\Omega,p,C)$ ensemble in $(D=4/d=3)$.  }
\label{fig:enf}
\end{figure}
The free energy $F$ behaves as a function of $\mathcal{T}$ for the various values of $(\varphi,\Omega,p,C)$. We plot the parametric plot $F(\mathcal{T})$ seen in figure \ref{fig:en6} and also its corresponding specific plot for its stability analysis in figure \ref{fig:enf}. We focus on figure \ref{fig:plotomegap} and \ref{fig:plotp} to study deeply the transitions and stability when plotting for various values of $\varphi$. We see for $\varphi=0.8,1$ (red, blue) we get two branches namely small (unstable)-large (stable) entropy branches which is (de)confinement phase transition and the transition point is where the plot crosses the free energy. In figure \ref{fig:plotp} for the same values we get Davies type phase transition seen for red and blue plots. For $\varphi=1.005$ there is a Van der Waals type phase transition where we get three branches of entropy analogous to black hole namely small (stable)-intermediate (unstable)-large (stable) entropy branches (purple). This type of branches continue to occur till we get to $\varphi=1.6, 1.8$ where the three branches converges to a single stable branch beyond this value showing only a single entropy phase. Now for plotting for various values of $\Omega$ is also seen in figure \ref{fig:plotomegap1},  \ref{fig:plotq} where for $\Omega=1.1,1.3$ we get two entropy branches namely small (unstable)-large(stable) undergoing (de)confinement phase transition in the $F(\mathcal{T})$ plot and Davies type phase transition in the specific heat plot. At $\Omega=1.42$ there is a Van der Waals transition where we see three branches namely small (unstable)-intermediate (stable) and a small unstable large branch (orange, purple). This goes on till $\Omega=1.46$ which is another critical point after which we get two unstable branches. The specific heat appears like a peak at that value and increases negatively till reaching a certain value of entropy then it decreases negatively and  it appears like an inverse lambda structure.

\section{Conclusion}
\label{sec:5}
In the beginning, we review very briefly the the mass-energy formulas for the extended phase space thermodynamics, then for the mixed thermodynamics, and lastly for the CFT thermodynamics for the Kerr-Newman AdS black hole comprising both the charge and rotating element \cite{dd,ii,dd1}. Then we investigated the various phase transitions for the dual holographic CFT states for the Kerr-Newman AdS black hole taking in various ensembles. Note that we were not able to take the entire ensemble as the interplay between the $(\mathcal{J},\Omega)$ and $(\mathcal{Q},\varphi)$ is quite complex, so we could only investigate a few ensembles. 
In the ensemble $(\mathcal{Q},\mathcal{J},\mathcal{V},C)$ in section \ref{sec:en1} we have plotted the free energy $F$ for the temperature parametrically with the entropy $\mathcal{S}$.
 We have plot three different plots seen in Figure \ref{fig:en1} that we plot for various values of electric charge $\mathcal{Q}$ kepping rest constant. For this, we have observed a first-order Van der Waals type phase transition for the range of $\mathcal{Q}<\mathcal{Q}_{crit}$. For when $\mathcal{Q}=\mathcal{Q}_{crit}$, we observe a kink-like structure which is described as a second-order superfluid $\lambda$ phase transition. At last, for $\mathcal{Q}>\mathcal{Q}_{crit}$ we see a smooth monotonous curve and there is no swallowtail or a kink anywhere. Now when we plot various values of $\mathcal{J}$, we get a first-order Van der Waals phase transition which is represented by a swallowtail for the range of angular momentum when $\mathcal{J}<\mathcal{J}_{crit}$ and then a second order superfluid $\lambda$ phase transition for the value when $\mathcal{J}=\mathcal{J}_{crit}$ and it is visualized as a kink like structure and then we observe a smooth curve for the value of $\mathcal{J}$ when $\mathcal{J}>\mathcal{J}_{crit}$. Lastly, for this ensemble we plot for the various values of the central charge $C$. When $C>C_{crit}$ we get the swallow tail-like structure which is the first-order Van der Waals phase transition at the point when the vertical line cuts the horizontal line. For $C=C_{crit}$, we get the second order superfluid $\lambda$ phase transition which is seen as a kink, and then a smooth monotonous curve for the central charge range that is $C<C_{crit}$\\
Now for the ensemble $(\mathcal{Q},\Omega,\mathcal{V},C)$ in section \ref{sec:en2}, we also have plotted the $F(\mathcal{T})$ graph for various values and have obtained three plots seen in Figure \ref{fig:en2}. Firstly we have plotted for various values of $\Omega$, keeping $\mathcal{Q}$, $\mathcal{V}$ and $C$ as constant. For this we get a swallowtail-like structure when the value of $\Omega$ is $\Omega<\Omega_{crit}$. This is also called the first-order Van der Waals type phase transition. Now when $\Omega=\Omega_{crit}$ we get a kink-like structure that corresponds to the second-order superfluid $\lambda$ phase transition and then for $\Omega>\Omega_{crit}$ we get a smooth monotonous curve. Now when we plot for the various values of $\mathcal{Q}$, we see a first-order Van der Waals phase transition for when $\mathcal{Q}<\mathcal{Q}_{crit}$ which is seen as a swallow tail structure. Then when $\mathcal{Q}=\mathcal{Q}_{crit}$ we see a kink-like structure which is second order superfluid $\lambda$ phase transition and then when $\mathcal{Q}>\mathcal{Q}_{crit}$ we get a smooth curve like structure. Now for this ensemble if we plot the curve for various values of the central charge $C$ for constant values of $\mathcal{Q}$, $\Omega$, and $\mathcal{V}$. We get the first order phase transition for $C>C_{crit}$ which we see a swallowtail behavior. Then for $C=C_{crit}$, we get the second-order superfluid $\lambda$ phase transition which also visually looks like a kink. and then lastly for $C<C_{crit}$ we get a smooth monotonous curve. \\
In the ensemble $(\varphi,\Omega,\mathcal{V},C)$ of section \ref{sec:en3}, we have plotted the $F(\mathcal{T})$ plot parametrically with entropy. When we plot the various values of  $\varphi$ while keeping the other parameters $\Omega$, $\mathcal{V}$ and $C$ as constant. We notice that for  $\varphi>\varphi_{crit}$, the free energy plot is single-valued where $F \leq 0$ and the curve cuts the $F$ axis. Now for the value of $\varphi<\varphi_{crit}$, we notice that the free energy curve consists of an upper and lower branch . For this branch, it changes sign at $F=0$ which is the first order phase transition. When $F<0$ it is said to be deconfined and when $F>0$, it is said to be confined. At the turning point we also see a Davies type phase transition. Now when we plot for various values of $\Omega$, there is a single value behavior of the Gibbs free energy and then there we see a bifurcation point where two branches appear At this point we see a Davies type phase transition. When the temperature increases from zero of the CFT, the ensemble in the confined state is dominated until we get a turning point where the free energy changes sign. Here only the first-order (de)confinement phase transition takes place. We can see the same kind of analogy when we plot for various values of $\mathcal{V}$. We also note that the (de)confinement phase transition does not depend on the central charge as we get no point where we get a single phase so the phase transition means that $\Omega$ and $\varphi$ depend on the CFT volume.\\
Now, section \ref{sec:en4}, in the ensemble $(\varphi,\mathcal{J},\mathcal{V},C)$, we have parametrically plot the $F(\mathcal{T})$ curve with the help of the entropy $\mathcal{S}$. Now when we have plotted for various values of $\mathcal{J}$, while having certain parameters constant like $\varphi$, $\mathcal{V}$, $C$. Here for $\mathcal{J}<\mathcal{J}_{crit}$ we see a swallow tail-like structure which is the first order Van der Waals phase transition. Then we see a second-order superfluid $\lambda$ phase transition which looks like a kink and is at the value where $\mathcal{J}=\mathcal{J}_{crit}$. Then when the value of $\mathcal{J}>\mathcal{J}_{crit}$ we see a smooth curve-like structure. Now then for various values of $C$ keeping the $\varphi$, $\mathcal{V}$ and $\mathcal{J}$ as constant. Here we have seen a first-order Van der Waals  phase transition that is a swallowtail-like structure when the value of the central charge is $C>C_{crit}$. Then a second-order superfluid $\lambda$ phase transition at the kink when $C=C_{crit}$. and then a smooth monotonous curve when $C<C_{crit}$ where we see no phase transition. Now when we plot for various values of $\varphi$, we get the same plots when plotted for $\mathcal{J}$. \\
  We also plot for the various $p$ ensembles in section \ref{sec:p1} and \ref{sec:p2}. In \ref{sec:p1} we plot for various values of $\Omega$ while keeping the rest constant. For $\Omega <\Omega_{crit}$ we get the first order Van der Waals type phase transition. At $\Omega=\Omega_{crit}$ we get the second order superfluid $\lambda$ phase transition and for $\Omega>\Omega_{crit}$ we get only a single phase. The same kind of phase transitions are seen when we plot for various different values of $\mathcal{Q}$. But when we plot for various values of $C$ we see Van der Waals type phase transition for $C>C_{crit}$ and at $C=C_{crit}$ we get the superfluid $\lambda$ phase transition and for $C<C_{crit}$ we get a single phase. Things get a little interesting in section \ref{sec:p2} where we get a plethora of phase transition like for $\varphi=0.8,1$, we get Davies type phase transition and also the (de)confinement phase transition of first order. Then for $\varphi=1.005, 1.1, 1.3, 1.6$ we see first order Van der Waals type phase transition and for $\varphi=1.8$ we see a single phases. When we plot for various values of $\Omega$ we get Davies type and (de)confinement phase transition for $\Omega=1.1,1.3$, Van der Waals type transition for $\Omega=1.42,1.44$ and what looks like a reverse $\lambda$ transition where both the branches are unstable and transits from a small entropy branch to a large entropy branch seen in the specific heat plot in figure \ref{fig:plotq}.  \\
In conclusion, this study focused on understanding the phase transitions and critical phenomena in charged rotating AdS black holes by examining the effects of introducing $\mathcal{V}$ (CFT volume), $p$ (CFT pressure) and $C$ (central charge) as additional parameters in the thermodynamic phase space through the analysis of six specific ensembles. Notably, the inclusion of these parameters revealed critical behaviors that were absent in traditional thermodynamics or the extended phase space thermodynamic (EPST) formalism.  The phase transitions seen are given in the table below which also highlights the key results obtained in the manuscript.\\

\begin{table}[h]
\centering
\begin{tabular}{|c|c|}
\hline
\textbf{Ensembles} & \textbf{Phase transitions} \\ \hline
$(\mathcal{Q},\mathcal{J},\mathcal{V},C)$ & Van der Waal, Superfluid $\lambda$ \\ \hline
$(\mathcal{Q},\Omega,\mathcal{V},C)$ & Van der Waal, Superfluid $\lambda$, Davies \\ \hline
$(\varphi,\Omega,\mathcal{V},C)$ & (de)confinement, Davies \\ \hline
$ (\varphi,\mathcal{J},\mathcal{V},C)$ &  Van der Waal, Superfluid $\lambda$ \\ \hline
$(\mathcal{Q},\Omega,p,C)$ &  Van der Waal, Superfluid $\lambda$ \\ \hline
$(\varphi,\Omega,p,C)$ & Van der Waal, Davies \\ \hline
\end{tabular}
\caption{Phase transitions in different ensemble}
\label{table:one}
\end{table}
In future endeavors, one can also examine the charge and rotating black holes in a more generalized form where we can also include the $\mu$ containing ensembles. One can also study more in-depth for each ensemble and check their coexistence plots, and also the order of phase transitions. We can also extend the study to a different formalism, namely the RPST (restricted phase space thermodynamics). On a concluding remark, the order and type of phase transition greatly depends on the ensemble and the number of variables taken into consideration. we have seen adding an extra variable, may it be electric charge $\mathcal{Q}$ or angular momentum $\mathcal{J}$ a plethora of phase transition is seen and also seen in the $p$ containing ensembles which was not seen before which opens a broader spectrum of ways via which phase transition can be studied. 

\section{Acknowledgement}
I extend my sincere gratitude to Mr. Bidyut Hazarika for his contribution while drafting the manuscript.

\section{Appendix}
\label{App}
\begin{equation}
A_1=\sqrt{\frac{A^3 \Lambda ^2}{G^3}-\frac{24 \pi  A^2 \Lambda }{G^2}+\frac{2304 \pi ^4 \left(G^2 Q^4+4 J^2\right)}{A}-\frac{48 \pi ^2 A \left(2 G^2 \Lambda  Q^2-3\right)}{G^2}+384 \pi ^3 \left(3 Q^2-2 J^2 \Lambda \right)}
\end{equation}

\begin{equation}
\resizebox{1\linewidth}{!}{$
A_2=\sqrt{3 \pi ^{5/2} \sqrt{\frac{1}{C P}} \left(32 J^2 P S+3 Q^4\right)+\frac{96 \pi ^{3/2} \left(4 \pi ^2 J^2+S^2\right)}{\sqrt{\frac{1}{C P}}}+\frac{2 \sqrt{6} S^2 \left(4 P S^2+3 \pi  Q^2\right)}{C}+48 \sqrt{\pi } P S^3 \sqrt{\frac{1}{C P}}+24 \pi ^2 \sqrt{6} Q^2 S}$}
\end{equation}

\begin{equation}
A_3=\sqrt{\frac{(4 \pi  C+S) \left(4 \pi  C S+\pi ^2 Q^2+S^2\right)^2}{C S V \left(16 \pi ^2 C-S V \Omega ^2+4 \pi  S\right)}}
\end{equation}

\begin{equation}
\resizebox{1.1\linewidth}{!}{$
A_{35}=\frac{A_3 \left(32 \pi ^2 C^2 \left(2 \pi ^3 Q^2+S^2 V \Omega ^2-10 \pi  S^2\right)-256 \pi ^4 C^3 S+4 \pi  C S \left(8 \pi ^3 Q^2+5 S^2 V \Omega ^2-28 \pi  S^2\right)+S^2 \left(\pi ^2 Q^2-3 S^2\right) \left(4 \pi -V \Omega ^2\right)\right)}{(4 \pi  C+S)^2 \left(4 \pi  C S+\pi ^2 Q^2+S^2\right)}$}
\end{equation}

\begin{equation}
A_{36}= \left(2 \pi ^3 Q^2+S^2 V \Omega ^2-10 \pi  S^2\right)+256 \pi ^4 C^3 S-4 \pi  C S \left(8 \pi ^3 Q^2+5 S^2 V \Omega ^2-28 \pi  S^2\right)
\end{equation}

\begin{equation}
\label{eq:en3}
\resizebox{1.\linewidth}{!}{$
\begin{split}
&A_4=-\frac{2 \sqrt{\mathcal{S}} \Omega  \sqrt{\frac{C \mathcal{S}^2 \varphi ^2 \mathcal{V}^2 \Omega ^2}{16 \pi ^2 C-\mathcal{S} \mathcal{V} \Omega ^2+4 \pi  \mathcal{S}}} \left(\frac{4 \pi  C \varphi ^2 \mathcal{V} (4 \pi  C+\mathcal{S})}{16 \pi ^2 C-\mathcal{S} \mathcal{V} \Omega ^2+4 \pi  \mathcal{S}}+4 \pi  C+\mathcal{S}\right)}{8 \pi ^{3/2} C \varphi  \sqrt{\mathcal{V}} \sqrt{4 \pi  C+\mathcal{S}}}-\frac{16 \pi  \sqrt{C} \sqrt{\mathcal{S}} \varphi ^2 \sqrt{\mathcal{V}} \sqrt{4 \pi  C+\mathcal{S}}}{8 \pi ^{3/2}\sqrt{16 \pi ^2 C-\mathcal{S} \mathcal{V} \Omega ^2+4 \pi  \mathcal{S}}}\\
&+ \frac{1}{\sqrt{\pi}}  B_1
+\frac{B_1\left(4 \pi  C \mathcal{S}^2 \left(\mathcal{V}^2 \Omega ^2 \left(\varphi ^2+2 \Omega ^2\right)-4 \pi  \mathcal{V} \left(\varphi ^2+9 \Omega ^2\right)+112 \pi ^2\right)+3 \mathcal{S}^3 \left(\mathcal{V} \Omega ^2-4 \pi \right)^2\right)}{8 \pi ^{3/2} (4 \pi  C+\mathcal{S})^2 \left(4 \pi  C \left(\varphi ^2 \mathcal{V}+4 \pi \right)+\mathcal{S} \left(4 \pi -\mathcal{V} \Omega ^2\right)\right)}-\\
&\frac{B_1\left(64 \pi ^3 C^2 \mathcal{S} \left(-2 \varphi ^2 \mathcal{V}-3 \mathcal{V} \Omega ^2+20 \pi \right)+256 \pi ^4 C^3 \left(4 \pi -\varphi ^2 \mathcal{V}\right)\right)}{8 \pi ^{3/2} (4 \pi  C+\mathcal{S})^2 \left(4 \pi  C \left(\varphi ^2 \mathcal{V}+4 \pi \right)+\mathcal{S} \left(4 \pi -\mathcal{V} \Omega ^2\right)\right)}
\end{split}
$}
\end{equation}

\begin{equation}
B_1=\sqrt{\frac{\mathcal{S} (4 \pi  C+\mathcal{S})^3 \left(4 \pi  C \left(\varphi ^2 \mathcal{V}+4 \pi \right)+\mathcal{S} \left(4 \pi -\mathcal{V} \Omega ^2\right)\right)^2}{C \mathcal{V} \left(16 \pi ^2 C-\mathcal{S} \mathcal{V} \Omega ^2+4 \pi  \mathcal{S}\right)^3}}
\end{equation}

\begin{equation}
\label{eq:temp3}
\resizebox{1\linewidth}{!}{$
\begin{split}
&A_5=-\frac{B_1\left(-128 \pi ^3 C^2 \mathcal{S} \varphi ^2 \mathcal{V}-256 \pi ^4 C^3 \varphi ^2 \mathcal{V}+8 \pi  C \mathcal{S}^2 \mathcal{V}^2 \Omega ^4+144 \pi ^2 C \mathcal{S}^2 \mathcal{V} \Omega ^2+3 \mathcal{S}^3 \mathcal{V}^2 \Omega ^4-24 \pi  \mathcal{S}^3 \mathcal{V} \Omega ^2\right)}{8 \pi ^{3/2} \mathcal{S} (4 \pi  C+\mathcal{S})^2 \left(4 \pi  C \varphi ^2 \mathcal{V}+16 \pi ^2 C-\mathcal{S} \mathcal{V} \Omega ^2+4 \pi  \mathcal{S}\right)}\\
&\frac{\left(-192 \pi ^3 C^2 \mathcal{S} \mathcal{V} \Omega ^2+1280 \pi ^4 C^2 \mathcal{S}+1024 \pi ^5 C^3+448 \pi ^3 C \mathcal{S}^2+48 \pi ^2 \mathcal{S}^3\right) B_1}{8 \pi ^{3/2} \mathcal{S} (4 \pi  C+\mathcal{S})^2 \left(4 \pi  C \varphi ^2 \mathcal{V}+16 \pi ^2 C-\mathcal{S} \mathcal{V} \Omega ^2+4 \pi  \mathcal{S}\right)}\\
&-\frac{B_1 \left(4 \pi  C \mathcal{S}^2 \varphi ^2 \mathcal{V}^2 \Omega ^2+16 \pi ^2 C \mathcal{S}^2 \varphi ^2 \mathcal{V}\right)}{8 \pi ^{3/2} \mathcal{S} (4 \pi  C+\mathcal{S})^2 \left(4 \pi  C \varphi ^2 \mathcal{V}+16 \pi ^2 C-\mathcal{S} \mathcal{V} \Omega ^2+4 \pi  \mathcal{S}\right)}
\end{split}
$}
\end{equation}

\begin{equation}
\label{eq:temp31}
\resizebox{1\linewidth}{!}{$
\begin{split}
&A_6=\frac{ \left(192 \pi ^3 \left(C \mathcal{V} \varphi ^2+4 C \pi \right)^3 \left(V \Omega ^2-4 \pi \right)^2\right) B_2}{\left(4 \pi -V \Omega ^2\right)^2 \left(16 \pi ^2 \left(C V \varphi ^2+4 C \pi \right) \left(4 C \pi  \left(V \varphi ^2+4 \pi \right)-4 \pi  \left(C V \varphi ^2+4 C \pi \right)\right) \left(4 C \pi -\frac{4 \pi  \left(C V \varphi ^2+4 C \pi \right)}{4 \pi -V \Omega ^2}\right)^2\right)}\\
&-\frac{ \left(64 C \pi ^3 \left(C V \varphi ^2+4 C \pi \right)^2 \left(V^2 \left(\varphi ^2+2 \Omega ^2\right) \Omega ^2+112 \pi ^2\right)\right) B_2}{\left(4 \pi -V \Omega ^2\right) \left(16 \pi ^2 \left(C V \varphi ^2+4 C \pi \right) \left(4 C \pi  \left(V \varphi ^2+4 \pi \right)-4 \pi  \left(C V \varphi ^2+4 C \pi \right)\right) \left(4 C \pi -\frac{4 \pi  \left(C V \varphi ^2+4 C \pi \right)}{4 \pi -V \Omega ^2}\right)^2\right)}\\
&-\frac{ \left(256 C^2 \pi ^4 \left(C V \varphi ^2+4 C \pi \right) \left(-2 V \varphi ^2-3 V \Omega ^2+20 \pi \right)\right) \left(256 C^3 \pi ^4 \left(4 \pi -V \varphi ^2\right)\right)B_2}{ \left(16 \pi ^2 \left(C V \varphi ^2+4 C \pi \right) \left(4 C \pi  \left(V \varphi ^2+4 \pi \right)-4 \pi  \left(C V \varphi ^2+4 C \pi \right)\right) \left(4 C \pi -\frac{4 \pi  \left(C V \varphi ^2+4 C \pi \right)}{4 \pi -V \Omega ^2}\right)^2\right)}\\
&\frac{\left(4 \pi -V \Omega ^2\right) B_2 \left(4 \pi  V \left(\varphi ^2+9 \Omega ^2\right)\right)}{16 \pi ^2 \left(C V \varphi ^2+4 C \pi \right) \left(4 C \pi  \left(V \varphi ^2+4 \pi \right)-4 \pi  \left(C V \varphi ^2+4 C \pi \right)\right) \left(4 C \pi -\frac{4 \pi  \left(C V \varphi ^2+4 C \pi \right)}{4 \pi -V \Omega ^2}\right)^2}
\end{split}
$}
\end{equation}

\begin{equation}
B_2=\sqrt{-\frac{\left(C V \varphi ^2+4 C \pi \right) \left(4 C \pi  \left(V \varphi ^2+4 \pi \right)-4 \pi  \left(C V \varphi ^2+4 C \pi \right)\right)^2 \left(4 C \pi -\frac{4 \pi  \left(C V \varphi ^2+4 C \pi \right)}{4 \pi -V \Omega ^2}\right)^3}{C V \left(4 \pi -V \Omega ^2\right) \left(\frac{4 \pi  V \left(C V \varphi ^2+4 C \pi \right) \Omega ^2}{4 \pi -V \Omega ^2}+16 C \pi ^2-\frac{16 \pi ^2 \left(C V \varphi ^2+4 C \pi \right)}{4 \pi -V \Omega ^2}\right)^3}}
\end{equation}

\begin{equation}
D_1=\sqrt[3]{\mathcal{S}^6+3 C \left(\mathcal{V} \varphi ^2+4 \pi \right) \mathcal{S}^5+3 C^2 \left(\mathcal{V}^2 \varphi ^4 \mathcal{S}^4+16 \pi ^2 \mathcal{S}^4+8 \pi  \mathcal{V} \varphi ^2 \mathcal{S}^4+72 \mathcal{J}^2 \pi ^3 \mathcal{V} \varphi ^2 \mathcal{S}^2\right)+C^3 D_{12}+D_{13}}
\end{equation}
\begin{equation}
D_{12}=864 \pi ^4 \mathcal{J}^2 \mathcal{S} \varphi ^2 \mathcal{V}+\mathcal{S}^3 \varphi ^6 \mathcal{V}^3+12 \pi  \mathcal{S}^3 \varphi ^4 \mathcal{V}^2+48 \pi ^2 \mathcal{S}^3 \varphi ^2 \mathcal{V}+64 \pi ^3 \mathcal{S}^3
\end{equation}
\begin{equation}
D_{13}=12 \sqrt{3} \pi ^{3/2} \sqrt{C^2 \mathcal{J}^2 \mathcal{S}^2 \varphi ^2 \mathcal{V} (4 \pi  C+\mathcal{S}) \left(3 C^2 D_{35}+C^3 D_{36}+3 C \mathcal{S}^4 \left(\varphi ^2 \mathcal{V}+4 \pi \right)+\mathcal{S}^5\right)}
\end{equation}
\begin{equation}
D_2=C^2 \mathcal{S}^2 \left(\mathcal{V} \varphi ^2+4 \pi \right)^2+\left(\mathcal{S}^2-D_1\right)^2+C \mathcal{S} \left(\mathcal{V} \varphi ^2 \left(D_1+2 \mathcal{S}^2\right)-8 \pi  \left(D_1-\mathcal{S}^2\right)\right)
\end{equation}
\begin{equation}
\begin{split}
&D_3=9 \left(12 \pi ^{3/2} \sqrt{3} \sqrt{C^2 \mathcal{J}^2 \mathcal{S}^2 \varphi ^2 \mathcal{V} (4 \pi  C+\mathcal{S}) \left(3 C^2 D_{35}+C^3 D_{36}+3 C \mathcal{S}^4 \left(\varphi ^2 \mathcal{V}+4 \pi \right)+\mathcal{S}^5\right)}\right.\\
&\left.+3 C^2 \left(72 \pi ^3 \mathcal{J}^2 \mathcal{S}^2 \varphi ^2 \mathcal{V}+\mathcal{S}^4 \varphi ^4 \mathcal{V}^2+8 \pi  \mathcal{S}^4 \varphi ^2 \mathcal{V}+16 \pi ^2 \mathcal{S}^4\right)+C^3 \left(864 \pi ^4 \mathcal{J}^2 \mathcal{S} \varphi ^2 \mathcal{V}\right.\right.\\
&\left.\left.+\mathcal{S}^3 \varphi ^6 \mathcal{V}^3+12 \pi  \mathcal{S}^3 \varphi ^4 \mathcal{V}^2+48 \pi ^2 \mathcal{S}^3 \varphi ^2 \mathcal{V}+64 \pi ^3 \mathcal{S}^3\right)+3 C \mathcal{S}^5 \left(\varphi ^2 \mathcal{V}+4 \pi \right)+\mathcal{S}^6\right){}^{2/3}
\end{split}
\end{equation}
\begin{equation}
D_{35}=\left(\mathcal{S}^3 \mathcal{V}^2 \varphi ^4+8 \pi  \mathcal{S}^3 \mathcal{V} \varphi ^2+36 \mathcal{J}^2 \pi ^3 \mathcal{S} \mathcal{V} \varphi ^2+16 \pi ^2 \mathcal{S}^3\right)
\end{equation}
\begin{equation}
D_{36}=\left(\mathcal{S}^2 \mathcal{V}^3 \varphi ^6+12 \pi  \mathcal{S}^2 \mathcal{V}^2 \varphi ^4+432 \mathcal{J}^2 \pi ^4 \mathcal{V} \varphi ^2+48 \pi ^2 \mathcal{S}^2 \mathcal{V} \varphi ^2+64 \pi ^3 \mathcal{S}^2\right)
\end{equation}

\begin{equation}
E_1=\sqrt{\left(-16 \pi ^2 c-4 \pi  S\right)^2-4 S \Omega ^2 \left(\frac{E_6}{E_5}-\frac{E_5}{3 \sqrt[3]{2} c p^2 \pi  S}\right)}
\end{equation}
\begin{equation}
E_2=\sqrt{\frac{(4 \pi  c+S) \left(\pi ^2 Q^2+S^2+4 c \pi  S\right)^2 \Omega ^2}{c \left(16 \pi ^2 c+4 \pi  S-E_1\right) \left(16 \pi ^2 c+4 \pi  S+\frac{1}{2} \left(-16 \pi ^2 c-4 \pi  S+E_1\right)\right)}}
\end{equation}
\begin{equation}
E_3=\left(8 \pi ^3 Q^2-28 \pi  S^2+\frac{5}{2} S \left(16 \pi ^2 c+4 \pi  S-E_1\right)\right)
\end{equation}
\begin{equation}
E_4=S^2 \left(3 S^2-\pi ^2 Q^2\right) \left(4 \pi -\frac{16 \pi ^2 c+4 \pi  S-E_1}{2 S}\right)
\end{equation}

\begin{equation}
E_5=\sqrt[3]{-108 c^2 p^4 \pi ^4 S^9-2160 c^3 p^4 \pi ^5 S^8-17280 c^4 p^4 \pi ^6 S^7-216 c^2 p^4 \pi ^6 Q^2 S^7-69120 c^5 p^4 \pi ^7 S^6-E_{51}+E_7}
\end{equation}
\begin{equation}
\begin{split}
&E_{51}=3456 c^3 p^4 \pi ^7 Q^2 S^6-138240 c^6 p^4 \pi ^8 S^5-108 c^2 p^4 \pi ^8 Q^4 S^5-20736 c^4 p^4 \pi ^8 Q^2 S^5-110592 c^7 p^4 \pi ^9 S^4\\
&-1296 c^3 p^4 \pi ^9 Q^4 S^4-55296 c^5 p^4 \pi ^9 Q^2 S^4-5184 c^4 p^4 \pi ^{10} Q^4 S^3-55296 c^6 p^4 \pi ^{10} Q^2 S^3-6912 c^5 p^4 \pi ^{11} Q^4 S^2
\end{split}
\end{equation}
\begin{equation}
\begin{split}
&E_6=\sqrt[3]{2} \left(\Omega ^2 S^6+12 c \pi  \Omega ^2 S^5+48 c^2 \pi ^2 \Omega ^2 S^4+2 \pi ^2 Q^2 \Omega ^2 S^4+64 c^3 \pi ^3 \Omega ^2 S^3+16 c \pi ^3 Q^2 \Omega ^2 S^3\right.\\
&\left.+\pi ^4 Q^4 \Omega ^2 S^2+32 c^2 \pi ^4 Q^2 \Omega ^2 S^2+4 c \pi ^5 Q^4 \Omega ^2 S\right)
\end{split}
\end{equation}
\begin{equation}
\resizebox{1.1\linewidth}{!}{$
E_7=\sqrt{108 c^3 \pi ^3 S^3 \left(\Omega ^2 S^6+12 c \pi  \Omega ^2 S^5+48 c^2 \pi ^2 \Omega ^2 S^4+2 \pi ^2 Q^2 \Omega ^2 S^4+64 c^3 \pi ^3 \Omega ^2 S^3+16 c \pi ^3 Q^2 \Omega ^2 S^3+\pi ^4 Q^4 \Omega ^2 S^2+32 c^2 \pi ^4 Q^2 \Omega ^2 S^2+4 c \pi ^5 Q^4 \Omega ^2 S\right)^3 p^6+E_{52}}$}
\end{equation}
\begin{equation}
\begin{split}
&E_{52}=\left(-108 c^2 p^4 \pi ^4 S^9-2160 c^3 p^4 \pi ^5 S^8-17280 c^4 p^4 \pi ^6 S^7-216 c^2 p^4 \pi ^6 Q^2 S^7-69120 c^5 p^4 \pi ^7 S^6\right.\\
&\left.-3456 c^3 p^4 \pi ^7 Q^2 S^6-138240 c^6 p^4 \pi ^8 S^5-108 c^2 p^4 \pi ^8 Q^4 S^5-20736 c^4 p^4 \pi ^8 Q^2 S^5-110592 c^7 p^4 \pi ^9 S^4\right.\\
&\left.-1296 c^3 p^4 \pi ^9 Q^4 S^4-55296 c^5 p^4 \pi ^9 Q^2 S^4-5184 c^4 p^4 \pi ^{10} Q^4 S^3-55296 c^6 p^4 \pi ^{10} Q^2 S^3-6912 c^5 p^4 \pi ^{11} Q^4 S^2\right)^2
\end{split}
\end{equation}
\begin{equation}
\begin{split}
&F_1=\text{Root}\left[c^2 \Phi ^4 \Omega ^4 S^7-2 c^2 \Phi ^6 \Omega ^2 S^7+12 c^3 \pi  \Phi ^4 \Omega ^4 S^6-24 c^3 \pi  \Phi ^6 \Omega ^2 S^6+48 c^4 \pi ^2 \Phi ^4 \Omega ^4 S^5-96 c^4 \pi ^2 \Phi ^6 \Omega ^2 S^5\right.\\
&\left.+64 c^5 \pi ^3 \Phi ^4 \Omega ^4 S^4-128 c^5 \pi ^3 \Phi ^6 \Omega ^2 S^4+\left(c \pi ^5 S^2 \Omega ^2 \Phi ^6+64 c p^2 \pi ^9 \Phi ^2+16 p^2 \pi ^8 S \Phi ^2\right) \text{$\#$1}^6\right.\\
&\left. +\left(8 c p \pi ^6 S^2 \Phi ^5+32 c^2 p \pi ^7 S \Phi ^5-8 c p \pi ^6 S^2 \Omega ^2 \Phi ^3-32 c^2 p \pi ^7 S \Omega ^2 \Phi ^3\right) \text{$\#$1}^5+\left(2 c \pi ^3 S^4 \Omega ^2 \Phi ^6\right.\right.\\
&\left.\left.+6 c^2 \pi ^4 S^3 \Omega ^2 \Phi ^6-8 c^3 \pi ^5 S^2 \Omega ^2 \Phi ^6+c^2 \pi ^4 S^3 \Omega ^4 \Phi ^4+4 c^3 \pi ^5 S^2 \Omega ^4 \Phi ^4-1024 c^3 p^2 \pi ^9 \Phi ^2-64 c p^2 \pi ^7 S^2 \Phi ^2\right.\right.\\
&\left.\left.-512 c^2 p^2 \pi ^8 S \Phi ^2+1024 c^3 p^2 \pi ^9 \Omega ^2+64 c p^2 \pi ^7 S^2 \Omega ^2+512 c^2 p^2 \pi ^8 S \Omega ^2\right) \text{$\#$1}^4+\left(8 c p \pi ^4 S^4 \Phi ^5\right.\right.\\
&\left.\left.+48 c^2 p \pi ^5 S^3 \Phi ^5-256 c^4 p \pi ^7 S \Phi ^5-8 c p \pi ^4 S^4 \Omega ^2 \Phi ^3-48 c^2 p \pi ^5 S^3 \Omega ^2 \Phi ^3+256 c^4 p \pi ^7 S \Omega ^2 \Phi ^3\right) \text{$\#$1}^3\right.\\
&\left.+\left(c \pi  S^6 \Omega ^2 \Phi ^6+4 c^2 \pi ^2 S^5 \Omega ^2 \Phi ^6-16 c^3 \pi ^3 S^4 \Omega ^2 \Phi ^6-64 c^4 \pi ^4 S^3 \Omega ^2 \Phi ^6+2 c^2 \pi ^2 S^5 \Omega ^4 \Phi ^4+16 c^3 \pi ^3 S^4 \Omega ^4 \Phi ^4\right.\right.\\
&\left.\left.+32 c^4 \pi ^4 S^3 \Omega ^4 \Phi ^4\right) \text{$\#$1}^2+\left(-1024 p \pi ^6 S^2 \Phi ^5 c^5+1024 p \pi ^6 S^2 \Phi ^3 \Omega ^2 c^5-768 p \pi ^5 S^3 \Phi ^5 c^4\right.\right.\\
&\left.\left.+768 p \pi ^5 S^3 \Phi ^3 \Omega ^2 c^4-192 p \pi ^4 S^4 \Phi ^5 c^3+192 p \pi ^4 S^4 \Phi ^3 \Omega ^2 c^3-16 p \pi ^3 S^5 \Phi ^5 c^2+16 p \pi ^3 S^5 \Phi ^3 \Omega ^2 c^2\right) \text{$\#$1}\&,1\right]
\end{split}
\end{equation}
\begin{equation}
F_2=\sqrt{-\frac{p^2 (4 \pi  c+S) F_1^2 \left(S^2+4 c \pi  S+\pi ^2 F_1^2\right)}{c S \Phi  \left(S^3 \Phi ^3+4 c \pi  S^2 \Phi ^3+\pi ^2 S F_1^2 \Phi ^3-\sqrt{S^2 \left(S^2+4 c \pi  S+\pi ^2 F_1^2\right)^2 \Phi ^6+64 p^2 \pi ^6 (4 \pi  c+S)^2 F_1^2}\right)}}
\end{equation}

\begin{equation}
F_3=\left(S^3 \Phi ^3+4 c \pi  S^2 \Phi ^3+\pi ^2 S F_1^2 \Phi ^3+32 c p \pi ^4 F_1+8 p \pi ^3 S F_1-F_4\right)
\end{equation}

\begin{equation}
F_4=\sqrt{S^2 \left(S^2+4 c \pi  S+\pi ^2 F_1^2\right)^2 \Phi ^6+64 p^2 \pi ^6 (4 \pi  c+S)^2 F_1^2}
\end{equation}

\begin{equation}
F_5=\frac{4 c \pi  \left(S^2+4 c \pi  S+\pi ^2 F_1^2\right)^2 \left(-S^3 \Phi ^3-4 c \pi  S^2 \Phi ^3-\pi ^2 S F_1^2 \Phi ^3+32 c p \pi ^4 F_1+8 p \pi ^3 S F_1+F_4\right)}{(4 \pi  c+S) F_3}
\end{equation}

\begin{equation}
\begin{split}
&F_6=\left(3 \Phi ^3 S^6+32 c \pi  \Phi ^3 S^5+112 c^2 \pi ^2 \Phi ^3 S^4+2 \pi ^2 \Phi ^3 F_1^2 S^4+24 p \pi ^3 F_1 S^4+128 c^3 \pi ^3 \Phi ^3 S^3\right.\\
&\left.+16 c \pi ^3 \Phi ^3 F_1^2 S^3+192 c p \pi ^4 F_1 S^3-3 F_4 S^3-\pi ^4 \Phi ^3 F_1^4 S^2-8 p \pi ^5 F_1^3 S^2+32 c^2 \pi ^4 \Phi ^3 F_1^2 S^2\right.\\
&\left.+384 c^2 p \pi ^5 F_1 S^2-20 c \pi  F_4 S^2-96 c p \pi ^6 F_1^3 S-32 c^2 \pi ^2F_4 S+\pi ^2 F_1^2 F_4 S-256 c^2 p \pi ^7 F_1^3\right)
\end{split}
\end{equation}

\end{document}